\documentclass[10pt,journal,compsoc,doublecolumn]{IEEEtran}
\usepackage[T1]{fontenc}
\usepackage[cmex10]{amsmath}
\usepackage[latin9]{inputenc}
\usepackage{xcolor}
\usepackage{pdfcolmk}
\usepackage{units}
\usepackage{color}
\usepackage{setspace}

\usepackage{amssymb}
\usepackage{epsfig}
\usepackage{stackrel}
\usepackage{graphicx}
\usepackage{esint}
\usepackage{cite}

\usepackage{epstopdf}
\usepackage{algorithm} 
\usepackage{algorithmic} 
\usepackage{textcomp} 
\usepackage{array}

\begin{document}

\title{
\textsc{Cost Efficiency Optimization of 5G Wireless Backhaul Networks}
}
\vspace{0.1cm}

\author{\normalsize
Xiaohu Ge$^1$,~\IEEEmembership{Senior~Member,~IEEE,} Song Tu$^1$, Guoqiang Mao$^2$,~\IEEEmembership{Senior~Member,~IEEE}, Vincent K. N. Lau$^3$,~\IEEEmembership{Fellow,~IEEE}, Linghui Pan$^1$\\
\vspace{0.70cm}
\small{
$^1$School of Electronic Information and Communications\\
Huazhong University of Science and Technology, Wuhan 430074, Hubei, P. R. China.\\
Email: \{xhge, songtu, lhpan\}@mail.hust.edu.cn\\
\vspace{0.2cm}
$^2$School of Computing and Communications\\
The University of Technology Sydney, Australia.\\
Email: g.mao@ieee.org\\
\vspace{0.2cm}
$^3$ Department of Electronic and Computer Engineering\\
Hong Kong University of Science and Technology, Hong Kong.\\
Email: eeknlau@ee.ust.hk
}\\
\thanks{\small{ Submitted to IEEE transaction on Mobile Computing.}}
}

\maketitle

\begin{abstract}
The wireless backhaul network provides an attractive solution for the urban deployment of fifth generation (5G) wireless networks that enables future ultra dense small cell networks to meet the ever-increasing user demands. Optimal deployment and management of 5G wireless backhaul networks is an interesting and challenging issue. In this paper we propose the optimal gateways deployment and wireless backhaul route schemes to maximize the cost efficiency of 5G wireless backhaul networks. In generally, the changes of gateways deployment and wireless backhaul route are presented in different time scales. Specifically, the number and locations of gateways are optimized in the long time scale of 5G wireless backhaul networks. The wireless backhaul routings are optimized in the short time scale of 5G wireless backhaul networks considering the time-variant over wireless channels. Numerical results show the gateways and wireless backhaul route optimization significantly increases the cost efficiency of 5G wireless backhaul networks. Moreover, the cost efficiency of proposed optimization algorithm is better than that of conventional and most widely used shortest path (SP) and Bellman-Ford (BF) algorithms in 5G wireless backhaul networks.
\end{abstract}

\IEEEpeerreviewmaketitle

\section{Introduction}
With the exponentially increasing demand for wireless data traffic in recent years, it has become evident that traditional macro cellular networks can not handle gigabit-level data traffic in an economical and environmental friendly way \cite{R1}. The fifth generation (5G) small cell network, adopting massive multiple input multiple output (MIMO) and millimeter wave transmission technologies, is emerging as a promising solution \cite{R2}. In order to reduce the  cell coverage sharply to achieve a high spatial spectrum efficiency, a large number of small cells have to be deployed to achieve a seamless coverage of urban regions and form 5G ultra-dense cellular networks \cite{R3}. However, it is uneconomical and cost-prohibitive for every small cell to be connected via the fiber to the cell (FTTC). As a consequence, wireless backhaul network becomes an indispensable part of 5G ultra-dense small cell network solutions\cite{R4}. To promote the deployment of 5G wireless backhaul networks, the cost efficiency optimization of 5G wireless backhaul networks is an inevitable problem.

Considering the significance of 5G wireless backhaul networks, some studies were discussed in \cite{R5,R6,R7,R8,R9}. The differences compared with the conventional massive MIMO for radio access networks and the benefits of the wireless backhaul employing massive MIMO were discussed in \cite{R5}. In \cite{R6}, Zhang \emph{et al.} provided a state-of-the-art survey on large-scale (LS)-MIMO studies and proposed a joint group power allocation and pre-beamforming scheme to substantially improve the performance of LS-MIMO-based wireless backhaul for heterogeneous wireless networks. Based on the beam alignment technique using adaptive subspace sampling and hierarchical beam codebooks, the millimeter wave beamforming transmission technology was developed for both wireless backhaul and access in small cell networks \cite{R7}. Dat \emph{et al.} proposed and experimentally demonstrated a seamlessly converged radio-over-fiber (RoF) and millimeter-wave system at 90 GHz for high-speed wireless signal transmission \cite{R8}. An in-band solution, i.e., multiplexing backhaul and access on the same frequency band, was proposed to solve the backhaul and inter base station (BS) coordination challenges \cite{R9}. The above results confirmed the potential for employing millimeter wave transmission technologies in wireless backhaul networks.

Since millimeter wave transmission technologies employing 60 GHz and 70$\sim$80 GHz are usually used for line-of-sight (LOS) links in short ranges \cite{Wei15,Zheng15}, multi-hop transmissions is needed for long-range transmissions in wireless backhaul networks adopting millimeter wave transmission technologies. Connectivity is an important issue to make all the nodes in multi-hop networks interconnected and reachable \cite{Ta09}. Some studies involved with the connectivity of wireless backhaul networks were explored in \cite{R10,R11,R13,R14}. Aimed at the joint maximization of energy and spectrum efficiency in wireless backhaul networks, a user association scheme was developed for heterogeneous wireless network where small cells forward their traffic through backhaul links to neighboring small cells until it eventually reaches the core network \cite{R10}. Considering the backhaul channel conditions and the quality of service requirements, an optimal joint routing and backhaul link scheduling scheme was proposed for a dense small cell network using 60 GHz multi-hop backhaul links \cite{R11}. Utilizing dual connectivity establishment methods, a self-organized multi-hop backhaul establishment procedure was developed to support autonomous bidirectional beam alignments for heterogeneous wireless backhaul networks \cite{R13}. Extended from a graph theoretic clique idea, a new adaptive backhaul architecture was proposed in \cite{R14} which allows changes to the overall backhaul topology and each individual backhaul link can vary its frequency to meet traffic demand. To avoid the blockage or link failure in multi-hop wireless backhaul networks, a group of super-BSs was configured to robustly relay backhaul traffic and minimize the resource cost on gateways \cite{D}. When wireless backhaul networks are provided by multiple mobile network operators (MNOs), a framework was proposed to optimize the route of wireless backhaul traffic based on the wireless channel conditions and economic factors among different MNOs \cite{E}. Higher capacity and energy consumption are required to aggregate and forward the wireless traffic into the next hop for aggregation nodes closing to the core network, such as the gateways. Therefore, the backhaul capacity bottleneck exists at the single gateway. However, the deployment of multiple gateways in wireless backhaul networks has not been considered in \cite{R11,R13,R14,D,E}. In our previous work \cite{Ge15}, energy efficiency of small cell backhaul networks was studied. Furthermore, a basic wireless backhaul network architecture with multiple gateways configurations was proposed in \cite{R15}. How to optimize the number and positions of multiple gateways in 5G wireless backhaul networks is still an open issue. Besides, the joint optimization of the multiple gateways deployment and wireless backhaul links has not been investigated in 5G wireless backhaul networks. Moreover, the total cost efficiency optimization  for 5G wireless backhaul networks is surprisingly rare in the open literature.

Motivated by the above observations, in this paper we propose a two-scale cost efficiency optimization solution for 5G wireless backhaul networks. The contributions and novelties of this paper are summarized as follows.
\begin{enumerate}
\item In traditional network cost models, the gateway was fixed and the cost of wireless backhaul network was optimized by the interference management in wireless links \cite{RA}. To avoid the performance loss due to the single objective optimization, the multiple performance aspects were formulated and optimized with a variety of system constraints in the microwave-based wireless backhaul network \cite{RB,RC}. To balance the wireless backhaul traffic in multiple gateways, we propose a cost efficiency model for 5G wireless backhaul networks considering multiple gateways deployment and millimeter-wave MIMO channel conditions. Since the millimeter wave transmission distance is short, the connectivity probability and non-isolation probability (probability that all small cell are not isolated) of 5G wireless backhual networks is analyzed.
\item To optimize the cost efficiency of 5G wireless backhaul networks, a two-scale joint optimization solution has been proposed. In the long time scale, the number and positions of gateways are optimized by the long time optimization (LTO) algorithm. In the short time scale, the wireless backhaul routings are optimized by the maximum capacity spanning tree (MCST) algorithm.
\item Numerical results show that there exists an optimal number of gateways for maximizing the cost efficiency of 5G wireless backhaul networks and the proposed algorithms are better than the conventional and widely used shortest path (SP) algorithm.
\end{enumerate}

The rest of this paper is organized as follows. Section II describes the system model of 5G wireless backhaul networks. The cost efficiency of 5G wireless backhaul networks is formulated in Section III. Furthermore, a two-scale joint optimization solution is proposed for the cost efficiency optimization of 5G wireless backhaul networks in Section IV. Simulation analysis is presented in Section V. Finally, Section VI concludes this paper.

\section{System Model}
\begin{figure}[!t]
\centerline{\includegraphics[width=8cm, draft=false]{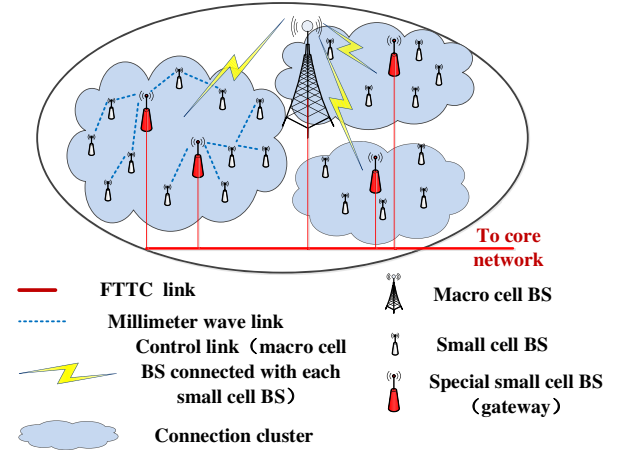}}
\caption{System model.}
\label{fig1}
\end{figure}

5G dense small cell networks equipped with massive MIMO antennas and millimeter wave transmission technology provide abundant resources, e.g. antennas and bandwidth, for wireless backhaul transmissions. In this paper, the user capacity requirements are assumed to be fully satisfied and the cost efficiency study therefore focuses on wireless backhaul networks. Considering the large path loss fading in millimeter wave propagations, the maximum distance of every hop in 5G wireless backhaul networks is limited to ${D_0}$ meters. The basic transmission models studied in this paper are described in Fig.~\ref{fig1}. To facilitate reading, the notations and symbols used in this paper are listed in Table 1.

\begin{table*}
\newcommand{\tabincell}[2]{\begin{tabular}{@{}#1@{}}#2\end{tabular}}
\centering
\caption{Notations and symbols}

\begin{tabular}{p{2.2cm}<{\centering}|p{15cm}}
\hline Notation &\hspace*{\stretch{1}}Description\hspace*{\stretch{1}} \\
\hline
${\cal V}$ &The set of $n$ SBSs \\
\hline
$B$ & The total number of connection clusters \\
\hline
$n$, $M$, $N$  &\tabincell{l}{The number of small cell BSs (SBSs), the number of SBSs configured as gateways, and the\\number of the rest non-gateway SBSs, respectively }    \\
\hline
$\chi$ &\tabincell{l}{ $\chi \in \Omega$ represents the spatial and temporal scheduling algorithm used in the wireless  backhaul \\network and $\Omega $ denotes the set of all scheduling algorithms} \\
\hline
$N_{i,j,T}^\chi $ &\tabincell{l}{ The number of bits transmitted by the SBS $SB{S_i}$ and which reached, i.e., successfully received\\ by, the respective  gateway $G{W_j}$ during a time interval $\left[ {0,T} \right]$}\\
\hline
${C^\chi }(M,N)$ &\tabincell{l}{The transport capacity of network using the spatial and temporal scheduling algorithm $\chi$ in a\\ connection cluster with $M$ gateways and $N$ SBSs }\\
\hline
$C(M,N)$ &\tabincell{l}{The transport capacity of network with $M$ gateways and $N$ SBSs in wireless backhaul network} \\
\hline
$\bar C(M,N)$ &\tabincell{l}{The average throughput of each gateway} \\
\hline
$\cal K$ &The set of all types of wireless traffic \\
\hline
${\cal L}$ & The set of all links for the wireless backhaul network \\
\hline
${a^\tau }$, $a_i^\tau $ &\tabincell{l}{The average transmission rate of the $\tau  - th$ wireless traffic and the $i-th$ SBS $SB{S_i}$ with the\\ $\tau  - th$ wireless traffic, respectively} \\
\hline
$r_{ji}^\tau $, $r_{iq}^\tau $ &\tabincell{c}{The incoming and outgoing transmission rates of the SBS $SB{S_i}$ with the $\tau  - th$ traffic type} \\
\hline
${d_\tau }$ & The destination for the $\tau  - th$ traffic \\
\hline
${\cal L}_i^{in}$, ${\cal L}_i^{out}$ &\tabincell{l}{ The set of input links and the set of output links at the SBS $SB{S_i}$, respectively}\\
\hline
${\cal V}_i^{in}$, ${\cal V}_i^{out}$ &\tabincell{l}{${\cal V}_i^{in} = \{ SB{S_j}:(SB{S_i},SB{S_j}) \in {\cal L}_i^{in}\} $, ${\cal V}_i^{out} = \{ SB{S_j}:(SB{S_i},SB{S_j}) \in {\cal L}_i^{out}\} $ , denote the set of\\ input SBSs and the set of output SBSs with respect to the SBS $SB{S_i}$, respectively} \\
\hline
${c_l}$ & The capacity of the link $l$   \\
\hline
$r_l^\tau $ & The transmission rate of the $\tau  - th$ traffic at the link $l$   \\
\hline
$\Psi$ &The large scale fading over the millimeter wave link\\
\hline
$\mathbf{H}$ & The wireless channel matrix of SBSs \\
\hline
$\eta$ & The number of path between transmitter and the receiver  \\
\hline
$\alpha _u$ & The small scale fading over the $u - th$ path  \\
\hline
$\theta _u^r$, $\theta _u^t$ &\tabincell{l}{The angle of arrival (AOA) and the angle of departure (AOD) for the $u - th$ path, respectively}\\
\hline
${{\bf{a}}_r}(\theta _u^r)$, ${{\bf{a}}_t}(\theta _u^t)$ & The receiving and transmitting antenna array response vectors, respectively\\
\hline
${{\bf{s}}_i}$ & The signal vector for the SBS $SB{S_i}$  \\
\hline
$N_{S,T}^i$ & The number of data streams at every SBS\\
\hline
${{\bf{x}}_i}$, ${{\bf{y}}_q}$ & The transmitted signal at the SBS $SB{S_i}$ and the  received signal at the SBS $SB{S_q}$  \\
\hline
${P_i}$ & The transmission power at the SBS $SB{S_i}$\\
\hline
${\bf{n}}$ & The additive white Gaussian noise (AWGN) with variance ${\sigma ^2}$\\
\hline
$N(\mathcal{A})$ & The number of SBSs in the special coverage of the MBS with the area $\mathcal{A}$\\
\hline
$D_0^{ni}$ &\tabincell{l}{The maximum value of the set of $\left\{ {{D_i}} \right\}$, where ${D_i}$ is the distance between a SBS\\ $SB{S_i},{\text{ }}1 \leqslant i \leqslant n$, and its closest SBS} \\
\hline
$D_0^{con}$ & \tabincell{l}{The longest link in the minimal spanning tree when all SBSs in the coverage of the MBS are\\ connected by a minimal spanning tree} \\
\hline
${W_i}$,${W_S}$ &\tabincell{l}{The transmission rate of wireless backhaul traffic at the SBS $SB{S_i}$ and the transmission rate of\\ backhaul traffic generated by a gateway, respectively} \\
\hline
${Y^\chi }(M,N)$ &The average number of transmissions for transmitting a bit to a gateway \\
\hline
${\Xi _j}$ & The set of SBSs associated with the gateway $G{W_j}$ \\
\hline
$t _{i,j,k,l}^\chi$ & The time required to transmit ${b_{i,j,k}}$ in the $l{\text{ - th}}$ transmission \\
\hline
$y_{\max }^\chi $ &\tabincell{l}{The maximum number of hops in all routes of wireless backhaul network with the spatial and\\ temporal scheduling algorithm $\chi$} \\
\hline
$\alpha $ & A small positive constant, independent of $T$\\
\hline
$e(M,N)$ & The cost efficiency of 5G wireless backhaul network with $M$ gateways and $N$ SBSs\\
\hline
${E_{EM}}$,${E_{OP}}$,${E_{G}}$ &\tabincell{l}{The total embodied energy and the total operation energy of wireless backhaul network, the\\ additional expense used for deploying the gateway, respectively} \\
\hline
$\bar W$& The average transmission rate of wireless backhaul traffic in the lifetime of the SBS\\
\hline
${P_{OP1}}$, ${P_{OP2}}$& The total operation energy consumed by gateways and SBSs, respectively\\
\hline
$\mathcal{N}$, $\mathcal{M}$ & The set of non-gateway SBSs and gateway SBSs, respectively \\
\hline
$S{E_l}$ & The spectrum efficiency over the link $l$\\
\hline
${B_s}$ & The bandwidth of link $l$\\
\hline
\end{tabular}
\label{tab1}
\end{table*}

\subsection{Connection Cluster Model}
The coverage of a macro cell BS (MBS) is assumed to be a circle with a radius $R$ and a total of $n$ small cell BSs (SBSs) are deployed in the coverage of the MBS. The MBS takes charge of the control plane and SBSs take charge of traffic transmission in this system. In this paper, the 5G wireless backhaul network comprises of SBSs in the coverage of a MBS. The distribution of SBSs is assumed to be governed by a Poisson point process with density $\mu$. Every SBS can connect with other SBSs within the distance ${D_0}$. The distance ${D_0}$ is the maximum transmission distance between two SBSs which is determined by the SBS transmission power. The set ${\cal V}$ includes $n$ SBSs and is divided into connection clusters. A group of SBSs is put into a connection culster if and only if these SBSs form a connected subnetwork. A pair of SBSs are connected if the distance between them is smaller than or equal to $D_0$. A set of SBSs form a connected subnetwork if and only if there is a path between any SBSs in the set to any other SBSs in the set. Let $B$ be the total number of connection clusters . There is no link between two connection clusters. The algorithm for forming connection clusters is given in Algorithm 1.

\begin{algorithm*}
\setcounter{algorithm}{0}
\renewcommand{\algorithmicrequire}{\textbf{Input:}}
\renewcommand\algorithmicensure {\textbf{Output:} }

\begin{algorithmic}
\caption{The generating algorithm of connection cluster.}

\REQUIRE The location of all the small cell BSs $\left\{ {({x_p},{y_p}),{SB{S_p}} \in {\cal V}}\right\}$, $n$\\
\begin{enumerate}
\STATE \textbf{Initialization:} The connection cluster ${\color{blue}\Upsilon}  = \emptyset$, the number of clusters is $B = 0$, the temporary cluster is $\Theta  = \emptyset $, $v = 0$, $t = 0$.
\item \textbf{while} $v < n$ \textbf{do} \\
\[\Theta  = \emptyset; v \leftarrow v + 1;t \leftarrow 0\]
\quad \textbf{if} {$B > 0$} \textbf{then}\\
\quad \quad \textbf{for} {$i = 1:B$} \textbf{do} \\
\quad \quad \quad \textbf{for} {$j = 1:\left| {{\Phi _i}} \right|$} \textbf{do} \\
\quad \quad \quad \quad The distance between small cell BS ${SB{S_v}}$ and ${SB{S_j}}$ is ${D_{vj}}$:
\[{D_{vj}} \leftarrow \sqrt {{{(x_v - x_j)}^2} + {{(y_v - y_j)}^2}} ;\] \\
\quad \quad \quad \quad \textbf{if} {${D_{vj}} \le {D_0}$} \textbf{then}\\
\[\Theta  = \Theta  + \left\{ {{SB{S_p}},{SB{S_p}} \in {\Phi _i}} \right\}; t \leftarrow t + 1; break;\]
\quad \quad \quad \quad \textbf{end if}\\
\quad \quad \quad \textbf{end for}\\
\quad \quad \textbf{end for}\\
\quad \textbf{end if}\\
\[B \leftarrow B + 1;\]
\quad \textbf{if} {$t =  = 0$} \textbf{then}\\
\[{\Phi _B} = \left\{ {SB{S_v}} \right\};\]
\quad \textbf{else}
\[{\Phi _B} = \left\{ {{SB{S_q}},{SB{S_q}} \in \Theta } \right\} + \left\{ {SB{S_v}} \right\};\]
\quad \quad \quad Delete the clusters which have been put into $\Phi _B$ and label the elements in {\color{blue}$\Upsilon$} in the original order.\\
\quad \textbf{end if}\\
\textbf{end while}\\
\end{enumerate}
\ENSURE Connection clusters ${\color{blue}\Upsilon}  = \left\{ {{\Phi _i}} \right\},i = 1,2,...B.$\\
\end{algorithmic}
\end{algorithm*}

Using the algorithm of forming connection clusters, $n$ SBSs are divided into $B$ connection clusters. To guarantee the forwarding of wireless backhaul traffic to the core network, every connection cluster must have at least one SBS configured as a gateway to connect with the core network. Therefore, the number of gateways must be larger than or equal to the number of connection clusters, i.e., $M \geqslant B$.

\subsection{Network Transport Capacity Model}
Without loss of generality, we assume that there are  $M$ SBSs out of the total $n$ SBSs configured as gateways, which connect to the core networks by FTTC links. It follows that the number of the rest non-gateway SBSs is equal to $N = n - M$. In this study we focus on the wireless backhaul traffic, i.e., the traffic transmitted from $N$ SBSs to the $M$ gateways, as the FTTC links connecting the gateway SBSs to the core network are considered to have ample bandwidth. Let $N_{i,j,T}^\chi $ be the number of bits transmitted by the SBS $SB{S_i}$ and which reached, i.e., successfully received by, the respective gateway $G{W_j}$ during a time interval $\left[ {0,T} \right]$ , with $T$ being an arbitrarily large number. The superscript $\chi  \in \Omega $ represents the spatial and temporal scheduling algorithm used in the wireless backhaul network and $\Omega $ denotes the set of all scheduling algorithms. We focus on the optimization of wireless backhaul networks. Thus, in this paper the spatial and temporal scheduling algorithm involves the selection strategy of backhaul gateways and wireless backhaul routings in 5G networks. If the same bit is transmitted from a SBS to multiple gateways, e.g., in the case of multicast, it is counted as one bit in the calculation of $N_{i,j,T}^\chi $.

It is assumed that the wireless backhaul network is stable. A wireless backhaul network is called stable if and only if the long-term incoming traffic rate into the wireless backhaul network equals the long-term outgoing traffic rate. It is further assumed that there is no traffic loss caused by queue overflow. The transport capacity of network using the spatial and temporal scheduling algorithm $\chi $ in a connection cluster with $M$ gateways and $N$ SBSs, denoted by ${C^\chi }(M,N)$, is defined as:
\[{C^\chi }(M,N) \buildrel \Delta \over = \mathop {\lim }\limits_{T \to \infty } \frac{{\sum\limits_{i = 1}^N {\sum\limits_{j = 1}^M {N_{i,j,T}^\chi } } }}{T}.\tag{1}\]
In this paper the transport capacity of network is focused on the wireless backhaul traffic in 5G dense small cell networks, which is calculated by the number of bits successfully transmitted among different SBSs \cite{Mao19}. Hence, the transport capacity of network with $M$ gateways and $N$ SBSs in the wireless backhaul network is defined by
\[C(M,N) \buildrel \Delta \over = \mathop {\max }\limits_{\chi  \in \Omega } {C^\chi }(M,N).\tag{2}\]
Considering $M$ gateways deployed in the wireless backhaul network, the average throughput of each gateway is given by
\[\bar C(M,N) \buildrel \Delta \over = \frac{{C(M,N)}}{M}.\tag{3}\]

\subsection{Link Traffic Model}
Assume that there exist $K$ types of wireless traffic in the wireless backhaul network. The set including all types of wireless traffic, such as video stream and voice traffic, is denoted as ${\cal K}: = \{ 1,...,\tau ,...,K\} $ and the set of all links is denoted as ${\cal L}$ for the wireless backhaul network. Without loss of generality, the  $\tau  - th$ type of wireless traffic is assumed to be transmitted by the link $l \in {\cal L}{\rm{(}}\tau {\rm{)}}$ and the average transmission rate of the $\tau  - th$ wireless traffic is assumed as ${a^\tau }$. The average transmission rate of the $i-th$ SBS $SB{S_i}$ with the $\tau  - th$ wireless traffic is denoted as $a_i^\tau $, which is used to evaluate the incoming traffic rate and outgoing traffic rate in a network. Hence, if the SBS $SB{S_i}$ is the traffic source of the $\tau  - th$ traffic then $a_i^\tau  = {a^\tau }$. If the SBS $SB{S_i}$ is the traffic destination of the $\tau  - th$ traffic then $a_i^\tau  =  - {a^\tau }$. If the SBS $SB{S_i}$ is the relaying SBS of the $\tau  - th$ traffic then $a_i^\tau  = 0$, which indicates the SBS  neither income the $\tau  - th$ traffic nor outgo the $\tau  - th$ traffic in the wireless backhaul network. The formulation of $a_i^\tau $ is expressed by
\[a_i^\tau  = \left\{ {\begin{array}{*{20}{l}}
{{a^\tau }{\rm{,  \ if \ }}SB{S_i}{\rm{ \ is \ source \ of \ }}\tau  - th{\rm{ \ traffic}}}\\
{ - {a^\tau }{\rm{,   \ if \ }}SB{S_i}{\rm{ \ is \ destination \ of \ }}\tau  - th{\rm{ \ traffic}}}\\
{0,{\rm{  \ otherwise}}}
\end{array}} \right..\tag{4}\]
The incoming and outgoing transmission rates of the SBS $SB{S_i}$ with the $\tau  - th$ traffic type are denoted as $r_{ji}^\tau $ and $r_{iq}^\tau $, respectively. When the SBS $SB{S_i}$ is configured as the source or relay SBS, the input traffic of the $\tau  - th$ traffic, including the incoming traffic and the generated traffic $a_i^\tau $, is equal to the output traffic of the $\tau  - th$ traffic at the SBS $SB{S_i}$ \cite{R18}, which is expressed by

\[a_i^\tau  + \sum\limits_{SB{S_j} \in {\cal V}_i^{in}} {r_{ji}^\tau }  = \sum\limits_{SB{S_q} \in {\cal V}_i^{out}} {r_{iq}^\tau } ,\]
\[{\rm{ }}\forall SB{S_i} \in {\cal V},SB{S_i} \ne {d_\tau },{\rm{ }} \tau  \in {\cal K},\tag{5}\]
\\
where ${d_\tau }$ is the destination for the $\tau  - th$ traffic, the set of input links and the set of output links at the SBS $SB{S_i}$ is denoted by ${\cal L}_i^{in}$ and ${\cal L}_i^{out}$, the set of input SBSs and the set of output SBSs with respect to the SBS $SB{S_i}$ are denoted by ${\cal V}_i^{in} = \{ SB{S_j}:(SB{S_i},SB{S_j}) \in {\cal L}_i^{in}\} $ and ${\cal V}_i^{out} = \{ SB{S_j}:(SB{S_i},SB{S_j}) \in {\cal L}_i^{out}\} $, respectively.

The capacity of the wireless link $l$ is denoted as ${c_l}$. The transmission rate of the $\tau  - th$ traffic at the link $l$ is denoted as $r_l^\tau $. When different types of wireless traffic are multiplexed over the same link $l$, the sum transmission rate of different types of wireless traffic should be less than or equal to the capacity of wireless link $l$ which is expressed as
\[\sum\limits_{\tau \in {\cal K}} {r_l^\tau }  \leqslant {c_l},{\rm{ }}l \in {\cal L}.\tag{6}\]

\subsection{Wireless Transmission Model}
Every SBS is equipped with ${N_T}$ and ${N_R}$ antennas for wireless transmission and reception, respectively. The millimeter wave frequency is adopted for wireless transmission in the wireless backhaul network. The large scale fading over the millimeter wave link is expressed by
\[\Psi {\rm{ = }}\beta {\rm{ + }}10\gamma {\rm{lo}}{{\rm{g}}_{10}}\Delta  + S,\tag{7a}\]
with
\[\beta  = 20{\rm{lo}}{{\rm{g}}_{10}}\left( {\frac{{4\pi }}{\lambda }} \right),\tag{7b}\]
where $\lambda $ is the wave length, $\gamma $ is the path loss coefficient, $\Delta $ is the distance between the transmitter and receiver, $S$ is the shadowing fading effect following a Gaussian distribution with zero mean and variance ${\xi ^2}$, i.e., $S \sim N\left( {0,{\xi ^2}} \right)$.

Assume that the millimeter wave transmission of SBSs is limited to line-of-sight (LOS) transmissions. The wireless channel matrix of SBSs is expressed by \cite{R19}
\[\begin{gathered}
  {\mathbf{H}} = \sqrt {\frac{{{N_T}{N_R}}}{{\Psi  \cdot \eta }}}  \cdot \sum\limits_{u = 1}^\eta  {{\alpha _u}{{\mathbf{a}}_r}(\theta _u^r){{\mathbf{a}}_t}{{(\theta _u^t)}^*}}  \hfill \\
  \ \ \  = \sqrt {\frac{{{N_T}{N_R}}}{{\Psi  \cdot \eta }}}  \cdot {{\mathbf{A}}_{\mathbf{R}}}{\mathbf{DA}}_{\mathbf{T}}^{\mathbf{*}}, \hfill \\
\end{gathered}\tag{8a}\]
with
\[{{\bf{A}}_{\bf{R}}} = \left[ {{{\bf{a}}_r}(\theta _1^r)|{{\bf{a}}_r}(\theta _2^r)|...|{{\bf{a}}_r}(\theta _\eta ^r)} \right],\tag{8b}\]
\[{{\bf{A}}_{\bf{T}}} = \left[ {{{\bf{a}}_t}(\theta _1^t)|{{\bf{a}}_t}(\theta _2^t)|...|{{\bf{a}}_t}(\theta _\eta ^t)} \right],\tag{8c}\]
\[{\bf{D}} = {\rm{diag}}\left\{ {{\alpha _1},{\alpha _2},...,{\alpha _\eta }} \right\},\tag{8d}\]
where $\eta $ is the number of paths between the transmitter and the receiver, ${\alpha _u}$ is the small scale fading over the $u - th$ path and is a complex normally distributed random variable with zero mean and unit variance, $\theta _u^r$ is the angle of arrival (AOA) and $\theta _u^t$ is the angle of departure (AOD) for the $u - th$ path, ${{\bf{a}}_r}(\theta _u^r)$ and ${{\bf{a}}_t}(\theta _u^t)$ are the receiving and transmitting antenna array response vectors, respectively. $\theta _u^r$ and $\theta _u^t$ are assumed to be uniformly distributed in the range of $[0,2\pi ]$ and then ${{\bf{a}}_r}(\theta _u^r)$ and ${{\bf{a}}_t}(\theta _u^t)$ are extended by \cite{R5}
\[{{\bf{a}}_r}(\theta _u^r) = \frac{1}{{\sqrt {{N_R}} }}{\left[ {1,{e^{j2\pi d\sin (\theta _u^r)/\lambda }},...,{e^{j2\pi ({N_R} - 1)d\sin (\theta _u^r)/\lambda }}} \right]^{\rm{T}}},\tag{9}\]
\[{{\bf{a}}_t}(\theta _u^t) = \frac{1}{{\sqrt {{N_T}} }}{\left[ {1,{e^{j2\pi d\sin (\theta _u^t)/\lambda }},...,{e^{j2\pi ({N_T} - 1)d\sin (\theta _u^t)/\lambda }}} \right]^{\rm{T}}},\tag{10}\]
where $d$ is the distance among antennas.
\begin{figure}[!t]
\centerline{\includegraphics[width=8cm, draft=false]{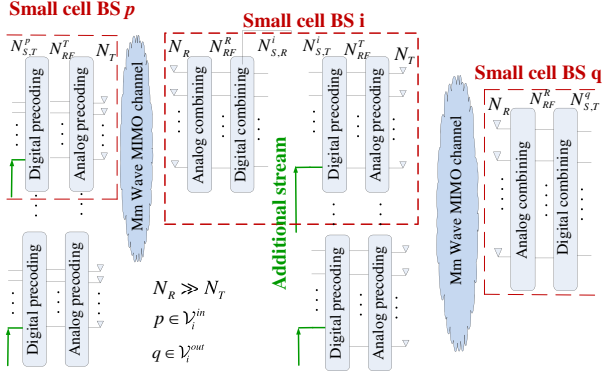}}
\caption{The millimeter wave MIMO transmission system.}
\label{fig2}
\end{figure}

The millimeter wave MIMO transmission system in the 5G wireless backhaul network is illustrated in Fig.~\ref{fig2}. For the multi-point to single-point transmission link $l$ in 5G wireless backhaul network, the transmitters include $Q$ SBSs each equipped with ${N_T}$ antennas and the receiver is a SBS equipped with ${N_R}$ antennas. Moreover, the number of antennas at receivers is assumed to be no less than $Q$ times of the antenna number at transmitters \cite{Miao2013}, i.e., ${N_R} \ge  Q{N_T}$, for spatial multiplexing. The numbers of radio frequency (RF) chains at the transmitter and receiver are $N_{RF}^T$ and $N_{RF}^R$, respectively. For the SBS $SB{S_i}$, the signal vector ${{\bf{s}}_i}$ consisting of $N_{S,T}^i$ data streams is processed by the digital precoding ${\bf{P}}_D^{} \in {^{N_{RF}^T \times N_{S,T}^i}}$ and then transmitted into $N_{RF}^T$ RF chains. Furthermore, the signal passed through RF chains is transmitted into ${N_T}$ transmission antennas by the analog precoding ${\bf{P}}_A^{} \in {^{{N_T} \times N_{RF}^T}}$. Hence, the transmitted signal at the SBS $SB{S_i}$ is expressed by ${{\bf{x}}_i} = {\bf{P}}_A^{}{\bf{P}}_D^{}{{\bf{s}}_i}$. When the digital and analog precoding methods is adopted for the beamforming in millimeter wave wireless transmissions, the interference from the wireless backhaul transmission of adjacent SBSs is assumed to be ignored in this paper. When ${{\bf{x}}_i}$ is received by the SBS $SB{S_q}$ with ${N_R}$ receive antennas, the received signal is expressed by
\[{{\bf{y}}_q} = \sqrt {{P_i}} {{\bf{H}}_l}{\bf{P}}_A^{}{\bf{P}}_D^{}{{\bf{s}}_i} + {\bf{n}},\tag{11}\]
where ${P_i}$ is the transmission power at the SBS $SB{S_i}$, ${\bf{n}}$ is the additive white Gaussian noise (AWGN) with variance ${\sigma ^2}$. Furthermore, the signal ${{\bf{y}}_q}$ is processed by the analog decoding ${\mathbf{F}}_A^{} \in {\mathbb{C}^{{N_R} \times N_{RF}^R}}$ and the digital decoding ${\mathbf{F}}_D^{} \in {\mathbb{C}^{N_{RF}^R \times N_{S,T}^i}}$. In the end, the received data streams are expressed by
\[{{\mathbf{\tilde y}}_q} = \sqrt {{P_i}} {\mathbf{F}}_D^*{\mathbf{F}}_A^*{{\mathbf{H}}_l}{\mathbf{P}}_A^{}{\mathbf{P}}_D^{}{{\mathbf{s}}_i} + {\mathbf{F}}_D^*{\mathbf{F}}_A^*{\mathbf{n}}.\tag{12}\]

\section{Cost Efficiency Formulation}
\subsection{Connectivity Probability and the Probability of Nodes Being Non-isolated}
The probability that there exist $n$ SBSs in a special coverage with area $\mathcal{A}$ is expressed by
\[{P_r}[n\;{\text{SBSs}}\;in\;\mathcal{A}] = {e^{ - \mu \mathcal{A}}}\frac{{{{(\mu \mathcal{A})}^n}}}{{n!}}.\tag{13}\]

\begin{figure}[!t]
\centerline{\includegraphics[width=8cm, draft=false]{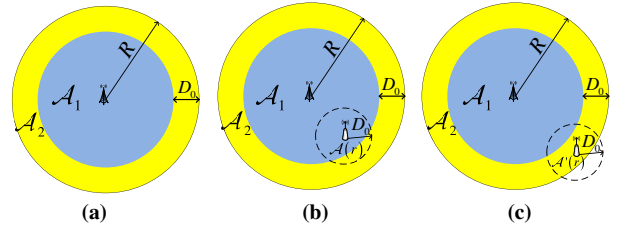}}
\caption{Coverage regions of the MBS and SBSs}
\label{fig3}
\end{figure}

Based on the system model in Fig.~\ref{fig1}, in this paper all SBSs are coveraged by a MBS. In this case, a SBS is isolated when the SBS can not establish a backhaul link with other SBSs in the given coverage of the same MBS. As shown in Fig.~\ref{fig3}(a), the coverage area of a MBS is divided into a circular region ${\mathcal{A}_1}$ with a radius $R - {D_0}$ (blue disk) and a annulus ${\mathcal{A}_2}$ with an inner radius of $R-D_0$ and an outer radius of $R$ (yellow ring). When a SBS is located in the circular region ${\mathcal{A}_1}$, the coverage area of the SBS in the coverage area of the MBS is $\mathcal{A}\left( r \right) = \pi D_0^2$, which is depicted in Fig.~\ref{fig3}(b). When a SBS is located in annulus ${\mathcal{A}_2}$, the coverage area of the SBS in the coverage area of the MBS is $\mathcal{A}'\left( r \right){\text{ = }}D_0^2 \cdot \arccos \tfrac{{{r^2} + D_0^2 - {R^2}}}{{2{D_0}r}} + {R^2} \cdot arccos\tfrac{{{r^2} - D_0^2 + {R^2}}}{{2Rr}} - \tfrac{1}{2}\sqrt \xi  ,{\text{ }}R - {D_0} \leqslant r \leqslant R$, with
$\xi  = (r + {D_0} + R)( - r + {D_0} + R)(r - {D_0} + R)(r + {D_0} - R)$, which is presented in Fig.~\ref{fig3}(c). Based on the illustration in Fig.~\ref{fig3}, the probability that a SBS is isolated is expressed by
\[\begin{array}{l}
P\left( {SBS \,  \text{is isolated}} \right) \\
= P\left( {SBS \,  \text{is isolated}\left| {SBS{\kern 1pt} {\kern 1pt} \text{is in}{\kern 1pt} {\kern 1pt} {{\cal A}_1}} \right.} \right)P\left( {SBS{\kern 1pt} {\kern 1pt} \text{is in}{\kern 1pt} {\kern 1pt} {{\cal A}_1}} \right)\\
{\rm{ + }}P\left( {SBS \, \text{is isolated}\left| {SBS{\kern 1pt} {\kern 1pt} \text{is in}{\kern 1pt} {\kern 1pt} {{\cal A}_2}} \right.} \right)P\left( {SBS{\kern 1pt} {\kern 1pt} \text{is in}{\kern 1pt} {\kern 1pt} {{\cal A}_2}} \right)\\
{\rm{ = }}{e^{ - \mu {\cal A}\left( r \right)}} \cdot \frac{{\pi {{\left( {R - {D_0}} \right)}^2}}}{{\pi {R^2}}} + \int_{R - {D_0}}^R {{e^{ - \mu {\cal A}'\left( r \right)}}}  \cdot \frac{{2\pi r \cdot dr}}{{\pi {R^2}}}.
\end{array}\tag{14}\]

Furthermore, the probability that there is no isolated SBS in the coverage of the MBS is expressed by
\[\begin{gathered}
  P(non-iso{\kern 1pt} {\kern 1pt} {\kern 1pt} SBS) = {(1 - P({SBS \,  is \, isolated}))^{E(N(\mathcal{A}))}} \hfill \\
  {\kern 1pt} {\kern 1pt} {\kern 1pt} {\kern 1pt} {\kern 1pt} {\kern 1pt} {\kern 1pt} {\kern 1pt} {\kern 1pt} {\kern 1pt} {\kern 1pt} {\kern 1pt} {\kern 1pt} {\kern 1pt} {\kern 1pt} {\kern 1pt} {\kern 1pt} {\kern 1pt} {\kern 1pt} {\kern 1pt} {\kern 1pt} {\kern 1pt} {\kern 1pt} {\kern 1pt} {\kern 1pt} {\kern 1pt} {\kern 1pt} {\kern 1pt} {\kern 1pt} {\kern 1pt} {\kern 1pt} {\kern 1pt} {\kern 1pt} {\kern 1pt} {\kern 1pt} {\kern 1pt} {\kern 1pt} {\kern 1pt} {\kern 1pt} {\kern 1pt} {\kern 1pt} {\kern 1pt} {\kern 1pt} {\kern 1pt} {\kern 1pt} {\kern 1pt} {\kern 1pt} {\kern 1pt} {\kern 1pt} {\kern 1pt} {\kern 1pt} {\kern 1pt} {\kern 1pt} {\kern 1pt} {\kern 1pt} {\kern 1pt} {\kern 1pt} {\kern 1pt} {\kern 1pt} {\kern 1pt} {\kern 1pt} {\kern 1pt} {\kern 1pt} {\kern 1pt} {\kern 1pt} {\kern 1pt} {\kern 1pt} {\kern 1pt} {\kern 1pt} {\kern 1pt} {\kern 1pt} {\kern 1pt} {\kern 1pt} {\kern 1pt} {\kern 1pt} {\kern 1pt} {\kern 1pt} {\kern 1pt} {\kern 1pt} {\kern 1pt} {\kern 1pt} {\kern 1pt} {\kern 1pt} = {(1 - P({SBS \,  is \, isolated}))^{\mu \pi {R^2}}}, \hfill \\
\end{gathered}\tag{15}\]
Where $E( \cdot )$ is the expectation operation, $N(\mathcal{A})$ is the number of SBSs in the special coverage of the MBS with the area $\mathcal{A}$.

\begin{figure}[!t]
\centerline{\includegraphics[width=8cm, draft=false]{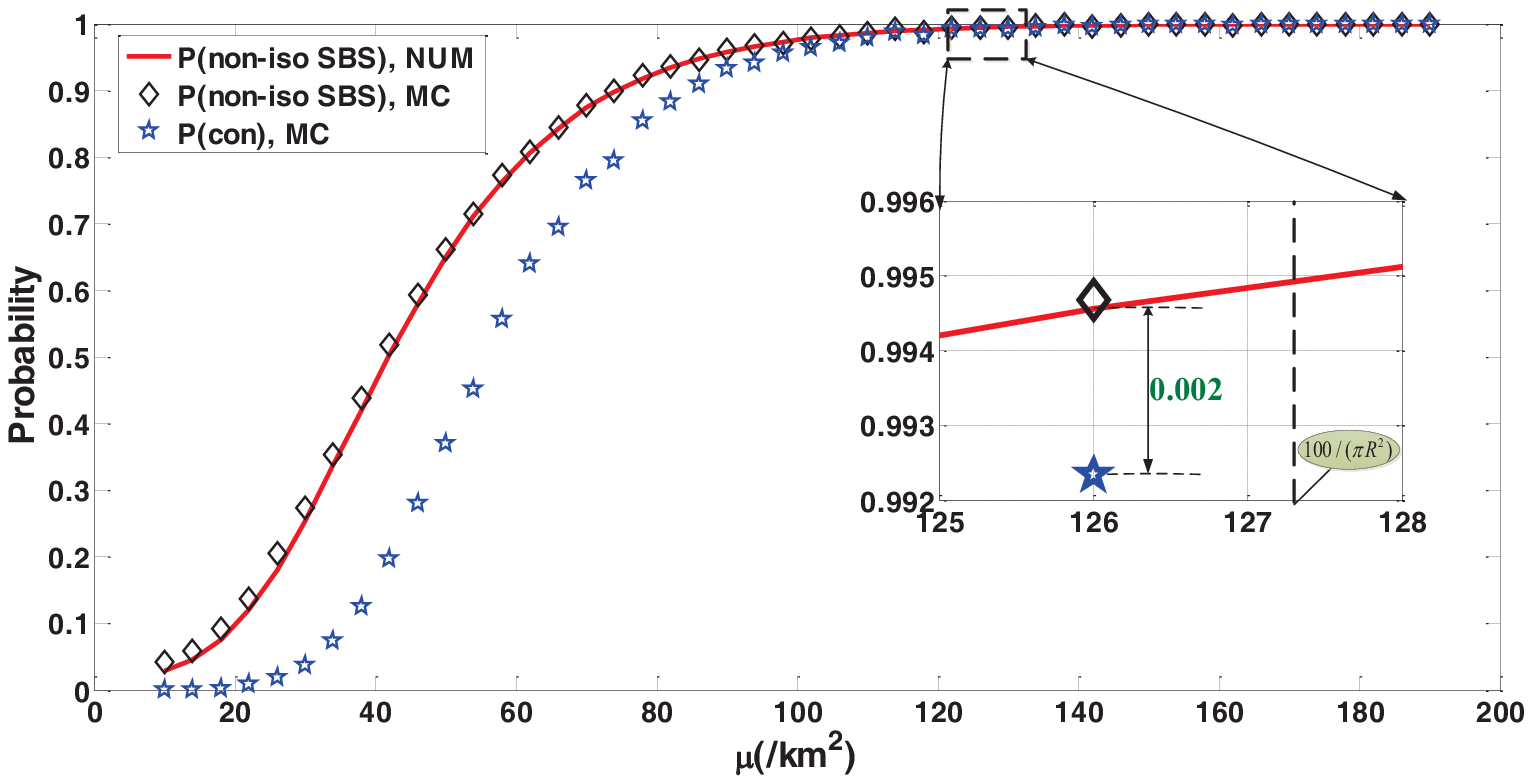}}
\caption{$P(non-iso{\kern 1pt} {\kern 1pt} {\kern 1pt} SBS)$ and $P(con)$}
\label{fig4}
\end{figure}

The probability that all SBSs are connected in the coverage of the MBS is denoted by $P(con)$. The event that there is no isolated SBS in the coverage of the MBS is the necessary condition for the event that all SBSs are connected in the coverage of the MBS. Hence, we can get a constrain as $P(con) \leqslant P(non-iso{\kern 1pt} {\kern 1pt} {\kern 1pt} SBS)$. To validate this constrain, $P(con)$ and $P(non-iso{\kern 1pt} {\kern 1pt} {\kern 1pt} SBS)$ are simulated by Monte Carlo (MC) and numerical (Num) methods in Fig.~\ref{fig4}, where the radius of coverage of the MBS is $R = 500$ meters and the radius of SBS is ${D_0} = 200$ meters.

From Fig.~\ref{fig4}, the numerical and the MC results of $P(non-iso{\kern 1pt} {\kern 1pt} {\kern 1pt} SBS)$ are coincident. This result implies that the expression of $P(non-iso{\kern 1pt} {\kern 1pt} {\kern 1pt} SBS)$ is reasonable. When the value of SBSs density $\mu $ is larger than or equal to 126, $P(no{\kern 1pt} n - iso{\kern 1pt} {\kern 1pt} {\kern 1pt} SBS)$ is approximated with $P(con)$. Therefore, we derive the following Theorem 1.\\
\textbf{Theorem 1}: When the coverage radius of the MBS $R$ and the coverage radius of SBSs ${D_0}$ are given, if the density of SBSs is large enough (for example, $100/\left( \pi {R^2} \right)$ can be one of the thresholds, as shown in Fig.~\ref{fig4}), the probability that there is no isolated SBSs in the coverage of the MBS and the probability that all SBSs are connected in the coverage of the MBS have the following relationships:$P(non-iso{\kern 1pt} {\kern 1pt} {\kern 1pt} SBS) - P(con) \to 0$, $P(non-iso{\kern 1pt} {\kern 1pt} {\kern 1pt} SBS) \to 1$ and $P(con) \to 1$.

\emph{Proof:}  When the density of SBSs distribution is configured as $\mu $, the distance between a SBS $SB{S_i},{\text{ }}1 \leqslant i \leqslant n$ and its closest SBS is ${D_i}$, the maximum value of the set of $\left\{ {{D_i}} \right\}$ is denoted as $D_0^{ni}$, which is the minimum value for satisfying the constraint that there is no isolated SBS in the coverage of the MBS. When all SBSs in the coverage of the MBS are connected by a minimal spanning tree, the longest link in the minimal spanning tree is denoted as $D_0^{con}$, which is the minimal value for satisfying the constraint that all SBSs are connected in the coverage of the MBS.

Based on results in \cite{Pen29}, when the density of SBSs in a square area is large enough, the value of $D_0^{ni}$ will approach to the value of $D_0^{con}$, i.e., $\mathop {\lim }\limits_{n \to \infty } P(D_0^{ni} = D_0^{con}) = 1$. When the density $\mu $ of SBSs in a circular area is large enough and the value of ${D_0}$ is fixed, we can derive a similar result, i.e., $P(non-iso{\kern 1pt} {\kern 1pt} {\kern 1pt} SBS) - P(con) \to 0$, $P(non-iso{\kern 1pt} {\kern 1pt} {\kern 1pt} SBS) \to 1$ and $P(con) \to 1$. Based on Theorem 1, we can use the value of $P(non-iso{\kern 1pt} {\kern 1pt} {\kern 1pt} SBS)$ to replace the value of $P(con)$ in the following simulation analysis when the density of SBS in the coverage of the MBS is large enough, e.g., $100/\left( {\pi {R^2} } \right)$.
\subsection{Network Transport Capacity of Wireless Backhaul Networks}
The optimization of wireless backhaul network can be achieved by the optimization of every connection cluster in the wireless backhaul network. Therefore, we propose the Theorem 2 to define the network transport capacity of a connection cluster in the wireless backhaul network.\\
\textbf{Theorem 2}: In the wireless backhaul network, the network transport capacity of a connection cluster consisting of $M$ gateways and $N$ SBSs satisfies:

{\footnotesize{
\[\begin{gathered}
C(M,N) \triangleq \min \left\{ {\mathop {\max }\limits_{\chi  \in \Omega } \frac{{\sum\limits_{i = 1}^N {{W_i}} }}{{{Y^\chi }(M,N)}} + M \cdot {W_S},{\text{ }}M \cdot {W_G}} \right\}, M < n,
\end{gathered}\normalsize{\tag{16a}}\]
}}
\[{W_i} = {a_i} + \sum\limits_{SB{S_j} \in {\cal V}_i^{in}} {{r_{ji}}} . \normalsize{\tag{16b}}\]
\\
where ${W_i},{\text{ 1}} \leqslant i \leqslant N$ is the transmission rate of wireless backhaul traffic at the SBS $SB{S_i}$, which includes the generated traffic ${a_i}$ from the SBS $SB{S_i}$ and the incoming transmission rate $\sum\limits_{SB{S_j} \in {\cal V}_i^{in}} {{r_{ji}}}$ at the SBS $SB{S_i}$. The generated traffic ${a_i}$ from the SBS $SB{S_i}$ is calculated by $a_i = \sum\limits_{\tau \in {\cal K}} {a_i^\tau}$, where $\cal K$ is the set including all types of wireless traffic. $r_{ji}^\tau $ is the incoming transmission rates of the SBS $SB{S_i}$ with the $\tau  - th$ traffic, which is calculated by $r_{ji} = \sum\limits_{\tau \in {\cal K}} {r_{ji}^\tau}$. ${Y^\chi }(M,N)$ is the average number of transmissions for transmitting a bit to a gateway. ${W_S}$ is the transmission rate of backhaul traffic generated by a gateway, which is configured to be the same for every gateway. ${W_G}$ is the gateway maximum transmission rate of backhaul traffic which includes the forwarding rate of wireless backhaul traffic generated from other SBSs and the transmission rate of backhaul traffic generated by a gateway. The set of SBSs associated with the gateway $G{W_j}$, i.e., forwarding wireless backhaul traffic into the gateway $G{W_j}$, is denoted as ${\Xi _j}$. Considering the function of gateway in wireless backhaul network, the sum of the forwarding rate of wireless backhaul traffic generated from other SBSs and the transmission rate of backhaul traffic generated by a gateway should be less than or equal to the gateway maximum transmission rate, i.e., $\sum\limits_{SB{S_i} \in {\Xi _j}} {{W_i}}  + {W_S} \leqslant {W_G},\forall G{W_j} \in \cal V$.

\emph{Proof:} Let ${b_{i,j,k}}$ be the $k{\text{ - th}}$ bit transmitted from the SBS $SB{S_i}$ to its destination gateway $G{W_j}$, $h_{i,j,k}^\chi $ be the number of transmissions required to deliver ${b_{i,j,k}}$ to its destination gateway when the spatial and temporal scheduling algorithm $\chi  \in \Omega $ is adopted. The average transmission number for transmitting a bit into a gateway is derived by
\[{Y^\chi }(M,N) = \mathop {\lim }\limits_{T \to \infty } \frac{{\sum\limits_{i = 1}^N {\sum\limits_{j = 1}^M {\sum\limits_{k = 1}^{N_{i,j,T}^\chi } {h_{i,j,k}^\chi } } } }}{{\sum\limits_{i = 1}^N {\sum\limits_{j = 1}^M {N_{i,j,T}^\chi } } }}.\tag{17}\]

Considering the stability of wireless backhaul networks, the backhaul traffic at all SBSs are less than or equal to the backhaul traffic at all gateways. In this case, the backhaul traffic at all SBSs at a time slot is denoted by $\sum\limits_{i = 1}^N {{W_i}} $ in the wireless backhaul network. The average backhaul traffic of a SBS is denoted by $\frac{{\sum\limits_{i = 1}^N {{W_i}} }}{N}$ in the wireless backhaul network. Let $t _{i,j,k,l}^\chi {\text{ , 1}} \leqslant l \leqslant h_{i,j,k}^\chi $, be the time required to transmit ${b_{i,j,k}}$ in the $l{\text{ - th}}$ transmission in the wireless backhaul network and is derived by $t _{i,j,k,l}^\chi  = \frac{N}{{\sum\limits_{i = 1}^N {{W_i}} }}$.\\
Remark 1. The total transmission time is first considered as the amount of traffic transmitted, measured in bits, multiplied by the time required to transmit each bit, in the wireless backhaul network on the individual SBS. Moreover, the total transmission time in the wireless backhaul network can also be calculated on the network level by evaluating the number of simultaneous transmissions in the entire wireless backhaul network. Obviously, the two values of total transmission time considering at SBSs and network level must be equal. On the basis of this observation, the Theorem 2 can be established.

At time $T$, the total transmission time ${T_{total}}$ during $\left[ {0,T} \right]$ includes the transmission time ${T_{gate}}$ for backhaul traffic that has reached its gateway and the transmission time ${T_{norm}}$ for backhaul traffic still in the transit at SBSs, i.e., ${T_{total}} = {T_{gate}} + {T_{norm}}$. Moreover, the transmission time ${T_{gate}}$ is calculated by
\[\begin{gathered}
  {T_{gate}} = \sum\limits_{i = 1}^N {\sum\limits_{j = 1}^M {\sum\limits_{k = 1}^{N_{i,j,T}^\chi } {\sum\limits_{l = 1}^{h_{i,j,k}^\chi } {t _{i,j,k,l}^\chi } } } }  \\
  \quad\quad\quad \ = \frac{N}{{\sum\limits_{i = 1}^N {{W_i}} }} \cdot \sum\limits_{i = 1}^N {\sum\limits_{j = 1}^M {\sum\limits_{k = 1}^{N_{i,j,T}^\chi } {h_{i,j,k}^\chi } } }.  \\
\end{gathered} \tag{18}\]

Let $y_{\max }^\chi $ be the maximum number of hops in all routes of wireless backhaul network with the spatial and temporal scheduling algorithm $\chi $, obviously $y_{\max }^\chi  \leqslant N$. Since the wireless backhaul network is stable, there exists a small positive constant $\alpha $, independent of $T$, such that the total amount of backhaul traffic in transit is bounded by $\alpha N$. Hence
\[\begin{gathered}
  {T_{norm}} \leqslant y_{\max }^\chi \alpha N t_{i,j,k,l}^\chi  \\
   \quad \leqslant \alpha {N^2}t _{i,j,k,l}^\chi  \\
   = \frac{{\alpha {N^3}}}{{\sum\limits_{i = 1}^N {{W_i}} }}. \\
\end{gathered} \tag{19}\]

On the other hand, the total transmission time during $\left[ {0,T} \right]$ calculated on the network level equals ${T_{total}} = N \cdot T$. Therefore,
\[\frac{N}{{\sum\limits_{i = 1}^N {{W_i}} }} \cdot \sum\limits_{i = 1}^N {\sum\limits_{j = 1}^M {\sum\limits_{k = 1}^{N_{i,j,T}^\chi } {h_{i,j,k}^\chi } } }  + {T_{norm}} = N \cdot T.\tag{20}\]

When the time interval of $\left[ {0,T} \right]$ is sufficiently large and the wireless backhaul network is stable, the amount of traffic in transit is negligibly small compared with the amount of traffic that has already reached its gateway. Furthermore, we can obtain the following result:
\[\mathop {\lim }\limits_{T \to \infty } \frac{{\sum\limits_{i = 1}^N {\sum\limits_{j = 1}^M {\sum\limits_{k = 1}^{N_{i,j,T}^\chi } {h_{i,j,k}^\chi } } } }}{{\sum\limits_{i = 1}^N {{W_i}}  \cdot T}} = 1.\tag{21}\]

Based on (1), (17) and (21), the transport capacity using the spatial and temporal scheduling algorithm $\chi $ in a connection cluster with $M$ gateways and $N$ SBSs is derived by
\[{C^\chi }(M,N) = \frac{{\sum\limits_{i = 1}^N {{W_i}} }}{{{Y^\chi }(M,N)}} + M \cdot {W_S}.\tag{22}\]

Based on (2), the network transport capacity of a connection cluster consisted of $M$ gateways and $N$ SBSs is derived by

\[C(M,N) = \mathop {\max }\limits_{\chi  \in \Omega } \frac{{\sum\limits_{i = 1}^N {{W_i}} }}{{{Y^\chi }(M,N)}} + M \cdot {W_S}.\tag{23}\]

Considering the maximum forwarding capacity of $M$ gateways $M \cdot {W_G}$ and the stability of wireless backhaul networks, the network  transport capacity of a connection cluster consisted of $M$ gateways and $N$ SBSs satisfies:
\[\min \left\{ {\mathop {\max }\limits_{\chi  \in \Omega } \frac{{\sum\limits_{i = 1}^N {{W_i}} }}{{{Y^\chi }(M,N)}} + M \cdot {W_S},{\text{ }}M \cdot {W_G}} \right\}.\tag{24}\]
\subsection{Formulation and Decomposition of Cost Efficiency Optimization}
With the massive MIMO and millimeter wave communication technologies adopting at 5G SBSs, SBSs have enough transmission rates used for wireless backhaul transmissions. However, the cell size of SBSs is obviously reduced, e.g. the coverage radius of 50 meters. To guarantee the seamless coverage of 5G small cell networks, SBSs have to be deployed by an ultra-dense deployment solution. Hence, there exist a large number of SBSs in the 5G wireless backhaul network. Based on the Theorem 1, SBSs in a given coverage, e.g. the coverage of a macro cell, are formed into one connection cluster if the density of SBSs is larger than a specific threshold. Considering that SBSs are ultra-densely deployed in the 5G wireless backhaul network, all SBSs are assumed to be formed into one connection cluster in 5G wireless backhaul network. Based on the Theorem 2, the network  transport capacity of wireless backhaul network increases with the increase of the number of gateways. However, the cost of wireless backhaul network is improved with the increasing of the number of gateways. Hence, it is a key issue for telecommunication providers to optimize the total cost efficient of 5G wireless backhaul networks.

Based on the result of Theorem 2, the cost efficiency of 5G wireless backhaul network with $M$ gateways and $N$ SBSs is defined as:
\[e(M,N) \triangleq \frac{{C(M,N)}}{{\zeta \left( {{E_{EM}} + {E_{OP}}} \right) + M \cdot {E_G}}},\tag{25a}\]
\[{E_{OP}}{\text{ = }}({P_{OP1}} + {P_{OP2}}) \cdot {T_{Lifetime}},\tag{25b}\]
\[{P_{OP1}} = M \cdot (a \cdot {P_{Norm}} \cdot {{{W_G}} \mathord{\left/
 {\vphantom {{{W_G}} {{W_0}}}} \right.
 \kern-\nulldelimiterspace} {{W_0}}} + b),\tag{25c}\]

\[{P_{OP2}} = N \cdot (a \cdot {P_{Norm}} \cdot {{{\bar W}} \mathord{\left/
{\vphantom {{{\bar W}} {{W_0}}}} \right.
\kern-\nulldelimiterspace} {{W_0}}} + b),\tag{25d}\]
\\
 where ${E_{EM}}$ is the total embodied energy of wireless backhaul network which is fixed as the 20\% of whole energy consumption of wireless backhaul network in the lifetime \cite{R1}, ${E_{OP}}$ is the total operation energy of wireless backhaul network, which is calculated by the total operation energy consumed by gateways ${P_{OP1}}$ and the total operation energy consumed by SBSs ${P_{OP2}}$ in the lifetime of gateways and SBSs, $\bar W$ is the average transmission rate of wireless backhaul traffic in the lifetime of the SBS, which is a constant. ${P_{Norm}}$ is the normalized transmission power associated with the normalized transmission rate ${W_0}$ at the gateway and SBS, $a$ and $b$ are fixed coefficients for computing the operation energy consumption \cite{R1}, $\zeta $ is the conversion factor between the energy consumption and cost expense, ${E_G}$ is the additional expense used for deploying the gateway.

 Furthermore, the cost efficiency optimization of 5G wireless backhaul networks is formulated as

\[\begin{gathered}
  \mathop {{\text{max}}}\limits_{M,{W_i},\chi  \in \Omega } e(M,N) \hfill \\
  s.t.{\text{  }}\left( 1 \right){W_i} \leqslant {c_l},{\text{ }}l \in \mathcal{L}_i^{out},\forall SB{S_i} \in \mathcal{N} \hfill \\
  \quad \ {\text{      }}\left( 2 \right)a_i^\tau + \sum\limits_{SB{S_p} \in \mathcal{V}_i^{in}} {r_{pi}^\tau}  = \sum\limits_{SB{S_q} \in \mathcal{V}_i^{out}} {r_{iq}^\tau} ,\hfill \\
  \quad \quad \quad {\rm{ }}\forall SB{S_i} \in {\cal V},SB{S_i} \ne {d_\tau },{\rm{ }} \tau  \in {\cal K} \hfill \\
  \quad \ {\text{      }}\left( 3 \right){P_i} \leqslant {P_{\max }},{\text{ }}\forall SB{S_i} \in \mathcal{N}, \hfill \\
\end{gathered} \tag{26}\]
\\
where $\mathcal{N}$ is the set of non-gateway SBSs(and $\mathcal{M}$, appearing in the next formula, is the set of gateway SBSs), ${P_{\max }}$ is the maximum transmission power at the SBS. To keep the stable of wireless backhaul networks, the flow balance constraint, i.e., (5) must be satisfied for the transport capacity of network in evaluating the cost efficiency of 5G wireless backhaul networks.

To optimize the cost efficiency of wireless backhaul networks, the optimization of gateways and wireless backhaul routes need to be solved for wireless backhaul networks. In general, the optimization of gateways, including the configuration of the number and locations of gateways, can stay for a long time after the wireless backhaul network has been deployed. Hence, the optimization of gateways in wireless backhaul networks can be updated in a long time scale.

  When the number and locations of gateways are fixed, based on (25b), (25c) and (25d), ${W_G}$ and $\bar W$ can be configured as constants for the cost efficiency of 5G wireless backhaul networks in the long time scale. In this case, the energy consumption of wireless backhaul network is assumed to be changeless. Furthermore, the optimization of cost efficiency is simplified to the optimization of wireless backhaul transport capacity of network, which benefits from the optimization of wireless backhaul routes. On the other hand, in the short time scale, the wireless backhaul routes are changed considering that the wireless channel capacity over every hop of wireless backhaul network is time-varying. As a consequence, a wireless backhaul routing scheme $\chi $ should be optimized in the short time scale of wireless backhaul networks.

Based on the optimization requirements of wireless backhaul network in the long time and short time scales, a two-scale joint optimization solution is formulated as follows:
\[\mathop {{\text{max}}}\limits_{M,{W_i},\chi  \in \Omega } e(M,N)\left\{ {\begin{array}{*{20}{c}}
{\mathop {{\rm{max}}}\limits_M e(M,N),{\rm{\; in \; long \; time \; scale;}}}\\
{\mathop {{\rm{max}}}\limits_{{W_i},\chi  \in \Omega } {C^\chi }(M,N),{\rm{\; in \; short \; time \; scale.}}}
\end{array}} \right.\tag{27a}\]
\[\begin{array}{l}
\mathop {{\text{max}}}\limits_M e(M,N)\\
s.t.{\text{ }}{\cal M} \cup {\cal N} = {\cal V},{\rm{ }}{\cal M} \cap {\cal N} = \phi.
\end{array}\tag{27b}\]

\[\begin{gathered}
  \mathop {{\text{max}}}\limits_{{W_i},\chi  \in \Omega } {C^\chi }(M,N) \hfill \\
  s.t.{\text{ }}W_i  \leqslant c_{_l}^{},{\text{ }}\forall l \in \mathcal{L} \hfill \\
  \quad \ {\text{      }}a_i^\tau + \sum\limits_{SB{S_p} \in \mathcal{V}_i^{in}} {r_{pi}^\tau}  = \sum\limits_{SB{S_q} \in \mathcal{V}_i^{out}} {r_{iq}^\tau} ,\hfill \\
  \quad \quad {\rm{ }}\forall SB{S_i} \in {\cal V},SB{S_i} \ne {d_\tau },{\rm{ }} \tau  \in {\cal K} \hfill \\
  \quad \ {\text{      }}{P_i} \leqslant {P_{\max }},{\text{ }}\forall SB{S_i} \in \mathcal{N}, \hfill \\
\end{gathered} \tag{27c}\]
\\
where $\cup $ and $\cap$ are operations of union and intersection on two sets, respectively.

Moreover, the channel status information (CSI) is important for optimizing the wireless backhaul routes in wireless backhaul networks. Without loss of generality, the following assumptions of CSI is declared in this study:
\begin{enumerate}
\item Every SBS can obtain the local CSI which includes the CSI over every wireless channel associated with the local SBS;
\item The macro cell BS can obtain all CSI of wireless channels in the wireless backhaul network;
\end{enumerate}

The spectrum efficiency over the link $l$ is expressed by
\[\begin{array}{l}
S{E_l} = {\log _2}\left( {\left| {{{\bf{I}}_{N_{S,T}^i}}} \right.} \right.\\
\quad\quad\left. {\left. { + \frac{{{P_i}}}{{\sigma _{}^2N_{S,T}^i}}{\bf{F}}_{\rm{D}}^*{\bf{F}}_{\rm{A}}^*{\bf{H}}_l^{}{{\bf{P}}_{\rm{A}}}{{\bf{P}}_{\rm{D}}}{\bf{P}}_{\rm{D}}^*{\bf{P}}_{\rm{A}}^*{\bf{H}}_l^*{{\bf{F}}_{\rm{A}}}{{\bf{F}}_{\rm{D}}}} \right| }\right).
\end{array}\tag{28}\]
Here, the interference from other links is assumed to be ignored as \cite{R10}. Moreover, the transmission capacity of the link $l$ is derived by
\[\begin{array}{l}
c_{_l}^{} = {B_s}S{E_l}\\
\,\quad = {B_s}{\log _2}\left( {\left| {{{\bf{I}}_{N_{S,T}^i}}} \right.} \right.\\
\quad\left. {\left. { + \frac{{{P_i}}}{{{\sigma ^2}N_{S,T}^i}}{\bf{F}}_{\rm{D}}^*{\bf{F}}_{\rm{A}}^*{\bf{H}}_l^{}{{\bf{P}}_{\rm{A}}}{{\bf{P}}_{\rm{D}}}{\bf{P}}_{\rm{D}}^*{\bf{P}}_{\rm{A}}^*{\bf{H}}_l^*{{\bf{F}}_{\rm{A}}}{{\bf{F}}_{\rm{D}}}} \right|} \right),
\end{array}\tag{29}\]
where ${B_s}$ is the bandwidth of link $l$. Based on the precoding/decoding optimization algorithms in \cite{R23}, the maximum transmission capacity of the link $l$ can be achieved. As a consequence, the optimization of wireless backhaul route can be achieved by maximizing the wireless transmission capacity of every link in wireless backhaul networks.

\section{Optimization Solution of Wireless backhaul Networks}
In this section, we give two algorithms to solve the two-scale joint optimization solution of (27), respectively.
\subsection{Solution of Long time Scale Gateways Optimization}
When BSs including the MBSs and SBSs are deployed in 5G dense small cell networks, the number and locations of BSs are per-determined. In this study, we first select the number and locations of gateways from the determined SBSs for maximizing the cost efficiency of wireless backhaul networks in long time scale. For a connection cluster with $n$ SBSs, we propose a new algorithm to obtain  the optimal number and location of gateways when the locations $\left\{ {({x_i},{y_i}),{\text{ 1}} \leqslant i \leqslant n} \right\}$ of SBSs $SB{S_i}$ are known. To easily design the optimization algorithm in the long time scale, the transmission rate of wireless backhaul traffic at SBSs is configured as the maximum transmission rate $W$, i.e., ${W_i} = W,{\text{ }}i \in \mathcal{N}$. The network transport capacity of a connection cluster is simplified as
\[C(M,N) \triangleq \mathop {\max }\limits_{\chi  \in \Omega } \frac{{NW}}{{{Y^\chi }(M,N)}} + M \cdot {W_S}.\tag{30}\]

To optimize the cost efficiency of wireless backhaul networks $e(M,N)$, the network transport capacity of a connection cluster needs to be maximized. To achieve the maximum network transport capacity of a connection cluster, the optimal solution can be implemented by a complete traversal method, i.e., all combination of SBSs are considered to find the optimal gateways, which can be proposed in our previous studies \cite{R15}. Considering the system model in this paper, the computation complexity of optimal algorithm in \cite{R15} is O(nM+3). Thus, it will cost much time for a large number of iterations. Algorithm 2 proposed here is a more efficient traversal algorithm which configures an initial set of gateways and then traverses the rest of gateways to replace the initial gateways for maximizing the network transport capacity. Compared with the optimal algorithm in \cite{R15}, the proposed Algorithm 2 is the suboptimal solution to maximize the network transport capacity of a connection cluster. Algorithm 2 is developed as follows.

\begin{algorithm*}
\begin{spacing}{0.75}
\setcounter{algorithm}{1}
\renewcommand{\algorithmicrequire}{\textbf{Input:}}
\renewcommand\algorithmicensure {\textbf{Output:} }
\begin{algorithmic}
\caption{Long Time Optimization Solution. (Part I)}
\REQUIRE $MAX\_M$, $n$, the location of all the small cell BSs $\left\{ {({x_p},{y_p}),{SB{S_p}} \in {\cal V}}\right\}$\\
\ENSURE $M_{opt}$, ${\bf{P}}{{\bf{S}}_M}$.\\
\textbf{for} {$M = 1:MAX\_M$} \textbf{do} \\
                        \quad The minimum average hop number of wireless backhaul traffic in the $M$ gateways macro cell is:
                        \[\mathop {\min }\limits_{\chi  \in \Omega } {Y^\chi }(M,N)\leftarrow {\rm{UnknowGateway}}\left(\left\{ {({x_p},{y_p}),SB{S_p} \in {\cal V}}\right\}, M \right );\]
                        \quad The position of $M$ gateways are:
                        \[{\bf{P}}{{\bf{S}}_M} = \left\{ {({x_q},{y_q}),SB{S_q} \in {\Phi _G}} \right\};\]
\textbf{end for}\\
Choose $M$ making energy efficiency to be the biggest:
\[M_{opt} \leftarrow {\rm{arg}}\mathop {{\rm{ max}}}\limits_M {\rm{ }}e(M,N);\]
\STATE \textbf{function} UnknowGateway $\left(\left\{ {({x_p},{y_p}),SB{S_p} \in {\cal V}}\right\}, M \right )$\\
    \begin{enumerate}
        \STATE \textbf{Initialization:} Put all the small cell BSs into the set of small cell BS ${\Phi _S}$ and empty the set of gateway ${\Phi _G}$.\\
                         \item \textbf{while} {$\left| {{\Phi _G}} \right| < M$} \textbf{do}
                             \[Array = zeros\]
                         \quad \textbf{for} {$SB{S_i}: SB{S_i} \in {\Phi _S}$} \textbf{do} \\
                         \quad \quad Put small cell BS $SB{S_i}$ into set $\Phi _G$:
                         \[\Phi _G = \Phi _G + \left\{SB{S_i}\right\};\]\\                     \quad \quad Call function KnowGateway, then save the result returned by KnowGateway into an array $Array$:
                         \[Arra{y_i} \leftarrow {\rm{KnowGateway}}\left( {\left\{ {({x_p},{y_p}),SB{S_p} \in {\cal V}}\right\},\left| {{\Phi _G}} \right|,\left\{ {({x_q},{y_q}),SB{S_q} \in {\Phi _G}} \right\}} \right);\]
                         \quad \quad Remove the small cell BS $SB{S_i}$ out of set $\Phi _G$:
                         \[\Phi _G = \Phi _G - \left\{SB{S_i}\right\};\]\\
                         \quad \textbf{end for}\\
                         \quad Put the small cell BS $SB{S_i}$ making $Arra{y_i}$ to be the biggest into set $\Phi _G$, and remove it from the set of small cell BS ${\Phi _S}$:\\
                         \[k \leftarrow \arg \mathop {\min }\limits_i Arra{y_i}\]
                         \[{\Phi _G} = {\Phi _G} + \left\{SB{S_k}\right\}; {\Phi _S} = {\Phi _S} - \left\{SB{S_k}\right\};\]
                         \textbf{end while}
                         \item \textbf{if} {$M > 1$} \textbf{then}\\
                         \quad \textbf{for} {$SB{S_j}: SB{S_j} \in {\Phi _G}$} \textbf{do}
                         \[Array = zeros, Arra{y_j} \leftarrow {\rm{KnowGateway}}\left( {\left\{ {({x_p},{y_p}), SB{S_p}\in {\cal V}} \right\},\left| {{\Phi _G}} \right|,\left\{ {({x_q},{y_q}),SB{S_q} \in {\Phi _G}} \right\}} \right);\]

                         \quad \quad \textbf{for} {$SB{S_i}: SB{S_i} \in {\Phi _S}$} \textbf{do} \\
                         \quad \quad \quad Exchange $SB{S_j}$ with $SB{S_i}$ (Put $SB{S_j}$ into the set of small cell, and remove it out of the set of gateway; Put $SB{S_i}$ into the set of gateway, and remove it out of the set of small cell), then
                         \[Arra{y_i} \leftarrow {\rm{KnowGateway}}\left( {\left\{ {({x_p},{y_p}), SB{S_p}\in {\cal V}} \right\},\left| {{\Phi _G}} \right|,\left\{ {({x_q},{y_q}),SB{S_q} \in {\Phi _G}} \right\}} \right);\]
                         Exchange back $SB{S_j}$ with $SB{S_i}$;\\
                         \quad \quad \textbf{end for}
                         \[k \leftarrow \arg \mathop {\min }\limits_i Arra{y_i}\]
                         \[{\Phi _G} = {\Phi _G} + \left\{SB{S_k}\right\}; {\Phi _S} = {\Phi _S} - \left\{SB{S_k}\right\};\]
                         \quad \quad \textbf{if} {$k \ne j$} \textbf{then}\\
                         \quad \quad \quad Exchange $SB{S_j}$ with $SB{S_k}$;\\
                         \quad \quad \textbf{end if}\\
                         \quad \textbf{end for}\\
                         \textbf{end if}\\
                         \end{enumerate}
\textbf{end function}\\
\end{algorithmic}
\end{spacing}
\end{algorithm*}

\begin{algorithm*}
\setcounter{algorithm}{1}
\renewcommand{\algorithmicrequire}{\textbf{Input:}}
\renewcommand\algorithmicensure {\textbf{Output:} }

\begin{algorithmic}
\caption{Long Time Optimization Solution. (Part II)}
\STATE \textbf{function} KnowGateway($\left\{ {({x_p},{y_p}),{SB{S_p}} \in {\cal V}}\right\}$, $M$, $\left\{ {({x_q},{y_q}),{\rm{ }}{G{W_q}} \in {\cal M}} \right\}$)\\
    \begin{enumerate}
        \STATE \textbf{Initialization:} For all the small cell BSs, $state(i) = 0,{ {{SB{S_i}} \in {\cal V}- {\cal M}}}$; All the gateways are 0 hop BSs.\\
                         \item \textbf{for} {${G{W_j}}: {G{W_j}} \in {\cal M}$} \textbf{do} \\
                         \quad Empty all the sets ${\Phi _h}$, variable $h \leftarrow 0$; Put gateway ${G{W_j}}$ into ${\Phi _0}$:
                         \[{\Phi _h} = {\Phi _h} + \left\{ {G{W_j}} \right\};\]
                       \quad \textbf{while} {${\Phi _h}$} \textbf{do} \\
                         \quad \quad \textbf{for} {$SB{S_k} \in {\Phi _h}$} \textbf{do} \\
                         \quad \quad \quad \textbf{for} {${SB{S_i}} \in {\cal V} - {\cal M}$} \textbf{do} \\
                         \quad \quad \quad \quad \quad The distance between small cell BS ${SB{S_i}}$ and $SB{S_i}$ is ${{D}_{ik}}$ :
                             \[{D_{ik}} \leftarrow \sqrt {{{(x_i - x_k)}^2} + {{(y_i - y_k)}^2}} ;\] \\
                         \quad \quad \quad \quad \textbf{if} {$state(i) =  = 0$\&\&${D_{ik}} \le {D_0}$} \textbf{then}\\
                         \quad \quad \quad \quad \quad The minimum hop number between small cell BS $SB{S_i}$ and gateway $G{W_j}$ is $h + 1$, then put small cell BS $SB{S_i}$ into set ${\Phi _{h+1}}$ and change the state of small cell BS $SB{S_i}$ $state(i)$ into $1$:
                         \[hop(i,j) \leftarrow h + 1;{\rm{ }}{\Phi _{h + 1}} = {\Phi _{h + 1}} + \left\{ SB{S_i} \right\};\]
                         \quad \quad \quad \quad \textbf{end if}\\
                         \quad \quad \quad \textbf{end for}\\
                         \quad \quad \textbf{end for}\\
                         \quad \quad Search the small cell BSs whose minimum hop number backhauling to gateway $G{W_j}$ is $h+1$:
                         \[h \leftarrow h + 1;\]
                       \quad \textbf{end while}\\
                         \textbf{end for}
                         \item The routing link with minimum hop number is selected for relaying the wireless backhaul traffic between the small cell BS $SB{S_i}$ and the corresponding gateway:
                             \[hop(i) \leftarrow \mathop {\min }\limits_{G{W_j} \in {\cal M}} hop(i,j);\]
                         \item The minimum average hop number of wireless backhaul traffic in the macro cell is calculated by:
                             \[\mathop {\min }\limits_{\chi  \in \Omega } {Y^\chi }(M,N) \leftarrow \frac{{\sum\limits_{SB{S_i} \in {\cal V} - {\cal M}}{hop(i)} }}{\left| { {\cal V} - {\cal M}} \right|};\]
                         \item \textbf{return} $\mathop {\min }\limits_{\chi  \in \Omega } {Y^\chi }(M,N)$.\\
                          \end{enumerate}
\textbf{end function}
\end{algorithmic}
\end{algorithm*}

The complexity of Algorithm 2 is analyzed in the following. The core functions of Algorithm 2 include the functions of KnowGateway() and UnknowGateway(). The calculation space of functions of KnowGateway() and UnknowGateway() depends on the number of SBSs in the coverage of MBS. The functions of KnowGateway() and UnknowGateway() must be convergent when the coverage of MBS is limited. Hence the Algorithm 2 must be convergent. For the function of \emph{KnowGateway()} in the Algorithm 2, the worst case occurs when there is only one SBS in the range of one hop. In this case, the complexity of function \emph{KnowGateway()} is $O{\text{(}}{n^3}{\text{)}}$. Moreover, the function \emph{KnowGateway()} is called by the function \emph{UnknowGateway()} with $O\text{(}n \text{)}$ times. Furthermore, the total complexity of Algorithm 2 is $O{\text{(}}{n^4}{\text{)}}$.

\subsection{Solution of Short Time Scale Routes Optimization}
\begin{figure}[!t]
\centerline{\includegraphics[width=8cm, draft=false]{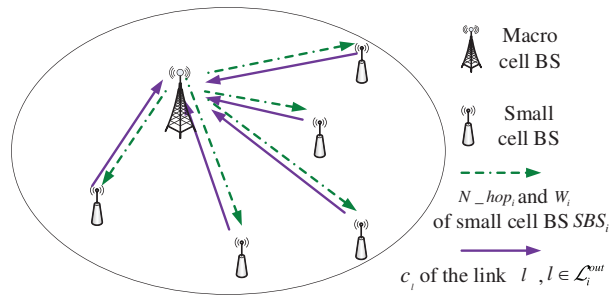}}
\caption{Wireless backhaul route schedule process}
\label{fig5}
\end{figure}

After the number and locations of gateways are optimized in the long time scale, wireless backhaul routes of wireless backhaul networks can be optimized in the short time scale. Based on system model in Fig.~\ref{fig1}, all SBSs report the local CSI to the macro cell BS in a time slot. The macro cell BS works out the optimal backhaul route information and then sends the optimal backhaul route information to all SBSs in the next time slot, as depicted in Fig.~\ref{fig5}.

Based on the CSI reported by $N$ SBSs, a directional connected graph with weight $G = \left( {\mathcal{V},\mathcal{L}} \right)$ is formed for the wireless backhaul network with $M$ gateways, where $\mathcal{V}$ is the set of SBSs and $\mathcal{L}$ is the link set of wireless backhaul routes. A SBS can have multiple input links but only  one output link in the weighted directional connected graph, where the weights of directed links correspond to the traffic rates. In the end, the wireless backhaul route has  a tree topology with the root node at a gateway. If there are multiple gateways in the wireless backhaul network, the wireless backhaul routes are represented by multiple tree topologies in the wireless backhaul network, i.e., every tree topology has a root node at a gateway. In this case, the short time scale optimization solution is obtained by first translating the weighted directed connected graph into multiple tree topologies. Moreover, the generated tree topology can maximize the network transport capacity of a connection cluster and satisfy three constrains: a) the root node of the tree topology is a gateway; b) the transmission rate of SBS wireless backhaul traffic, which corresponds to the weight of corresponding directed link, is less than or equal to the wireless channel capacity; c) the wireless backhaul traffic need to be balanced at SBSs. Furthermore, the average transmission number is calculated by

\[{Y^\chi }(M,N){\text{ = }}\frac{{\sum\limits_{i = 1}^N {ho{p_i}} }}{N},\tag{31}\]
where $ho{p_i}$ is the hop number between the SBS $SB{S_i}$ and the gateway. Thus, the network transport capacity of wireless backhaul network is calculated by

\[C(M,N) = \mathop {\max }\limits_{\chi  \in \Omega } \frac{{N \cdot \sum\limits_{i = 1}^N {{W_i}} }}{{\sum\limits_{i = 1}^N {ho{p_i}} }}.\tag{32}\]

Based on three constrains and (32), a maximum capacity spanning tree (MCST) algorithm is developed to obtain the tree topology with maximum network transport capacity $T = \left( {\mathcal{U},\mathcal{T}\mathcal{L}} \right)$, where $\mathcal{U}$ is the set of nodes, $\mathcal{T}\mathcal{L}$ is the link set. The detail MCST algorithm is illustrated in Algorithm 3. The calculation space of Algorithm 3 depends on the set   which includes all gateways in the coverage of MBS. The Algorithm 3 must be convergent when the number of gateways is limited in the coverage of MBS. The complexity of Algorithm 3 is O(n2). To optimization the cost efficiency of 5G wireless backhaul networks, the number and locations of gateways can be adjusted by Algorithm 2 per month/year and then the wireless backhaul routes can be adjusted by Algorithm 3 per hour/day.
\begin{algorithm*}
\setcounter{algorithm}{2}
\renewcommand{\algorithmicrequire}{\textbf{Input:}}
\renewcommand\algorithmicensure {\textbf{Output:} }

\begin{algorithmic}
\caption{MCST Algorithm.}

\REQUIRE The set of gateways ${\cal M}$ and the set of SBSs ${\cal N}$, wireless channel capacities over all wireless links in the wireless backhual netwok.\\
\begin{enumerate}
\STATE \textbf{Initialization:} The set of node ${\cal U} = {\cal M}$, the set of candidate links ${\cal C}{\cal L}$ including all links between nodes ${Z_j} \in {\cal M}, 1\leqslant j\leqslant M$ and ${Z_i} \in {\cal N}, 1\leqslant j\leqslant N$, the hop number between the gateway and the node ${Z_j} \in {\cal M}$ is 0, empty the set of link ${\cal T}{\cal L}$.
\item \textbf{while} {${\cal V} - {\cal U}$} \textbf{do} \\
\quad Traverse the link in the set of candidate links $\left(\left( {{Z_j},{Z_v}} \right),{\rm{ }}{Z_j} \in {\cal U},{Z_v} \in {\cal V} - {\cal U}\right)$, then choose the link making $\frac{{\left( {\left| {{\cal U}{\rm{ - }}{\cal M}} \right| + 1} \right) \cdot \left( {\sum\limits_{{Z_i} \in {\cal U}{\rm{ - }}{\cal M}} {{W_i}} {\rm{ + }}{W_{v,tmp}}} \right)}}{{\sum\limits_{{Z_i} \in {\cal U}{\rm{ - }}{\cal M}} {ho{p_i}}  + ho{p_{v,tmp}}}}$ to be the biggest capacity (where $\text{ }ho{p_{v,tmp}}\leftarrow ho{p_j} + 1$ is the number of hop between the gateway and a temporary node ${Z_{v,tmp}} \in {\cal V} - {\cal U}$, ${W_i}$ and ${W_{v,tmp}}$ are restricted by the constrains of b) and c)), then\\
\quad Put $\left( {{Z_j},{Z_v}} \right)$ into the set of links ${\cal T}{\cal L}$; Put ${Z_v}$ into ${\cal U}$:
\[{\cal U}{\rm{ = }}{\cal U}{\rm{ + \{ }}{Z_v}{\rm{\} }};\]
\quad The number of hop $ho{p_v}$ between the gateway and the node ${Z_v}$ is one more than the number $ho{p_j}$ of hop between the gateway and the node ${Z_j}$:
\[ho{p_v} \leftarrow ho{p_j} + 1;\]
\quad The next hop node $N\_ho{p_v}$ between the gateway and the node ${Z_v}$ is assigned by ${Z_j}$:
\[N\_ho{p_v}{\rm{ = }}{Z_j};\]
\quad Update the candidate links in ${\cal C}{\cal L}$: The set of candidate links ${\cal C}{\cal L}$ only includes all links between nodes ${Z_j} \in {\cal U}$ and ${Z_v} \in {\cal V}-{\cal U}$\\
\textbf{end while}\\
\end{enumerate}
\ENSURE ${\cal T}{\cal L}$, $\bf{N\_hop}$.\\
\end{algorithmic}
\end{algorithm*}

\section{Simulation Results and Discussion}
Based on the proposed two-scale joint optimization algorithm, the effect of various system parameters on the cost efficiency and transport capacity of network is analyzed and compared by numerical simulations in this section. Without loss of generality, the number of data streams $N_{S,T}^i$ at every SBS is configured to be the same. The maximum transmission rate of wireless backhaul traffic at every SBS is configured as $W = 10{\text{Gbps}}$ considering the millimeter wave technology. The detailed parameters of simulation system are illustrated in Table~\ref{tab1}.

\begin{table*}
\centering
\caption{Default parameters of simulation systems}
\begin{tabular}{l|c}
\hline Parameters & Default values \\
\hline
The maximum distance of every hop ${D_0}$ & 200 meters \cite{R31} \\
\hline
The radius of macro cell $R$ & 500 meters \\
\hline
The density of SBSs in a wireless backhaul network $\mu $ & 100$/\left( {\pi {R^2} \cdot k{m^2}} \right)$ \\
\hline
The conversion factor $\zeta $ & 1 \texteuro/kWh \cite{R24} \\
\hline
Lifetime of SBS ${T_{Lifetime}}$ & 5 years \cite{R30} \\
\hline
The fixed coefficient $a$ & 7.84 \\
\hline
The fixed coefficient $b$ & 71.5 Watt \\
\hline
The normalized transmission power at SBS ${P_{Norm}}$ & 1 Watt \\
\hline
The normalized transmission rate at SBS ${W_0}$ & 1 Gbps \\
\hline
The additional expense used for deploying the gateway $E$ & 3900\texteuro \cite{R25} \\
\hline
Wave length of millimeter wave $\lambda $ & 5 millimeters \\
\hline
Path loss coefficient $\gamma $ & 2 \\
\hline
Distance among antennas $d$ & 2.5 millimeters \\
\hline
Number of transmission antennas at SBS ${N_T}$ & 16 \\
\hline
Number of receive antennas at SBS ${N_R}$ & 128 \\
\hline
Number of RF chains at a transmitter $N_{RF}^T$ & 4 \\
\hline
Number of RF chains at a receiver $N_{RF}^R$ & 4 \\
\hline
Number of data stream $N_{S,T}^i$ & 2 \\
\hline
The maximum transmission power ${P_{\max }}$ & 1 Watt \cite{R26} \\
\hline
The maximum forwarding capacity of the gateway ${W_G}$ & 100 Gbps  \\
\hline
Bandwidth of SBSs ${B_s}$ & 1 GHz \\
\hline
\end{tabular}
\label{tab1}
\end{table*}

\begin{figure}[!t]
\centerline{\includegraphics[width=9cm, draft=false]{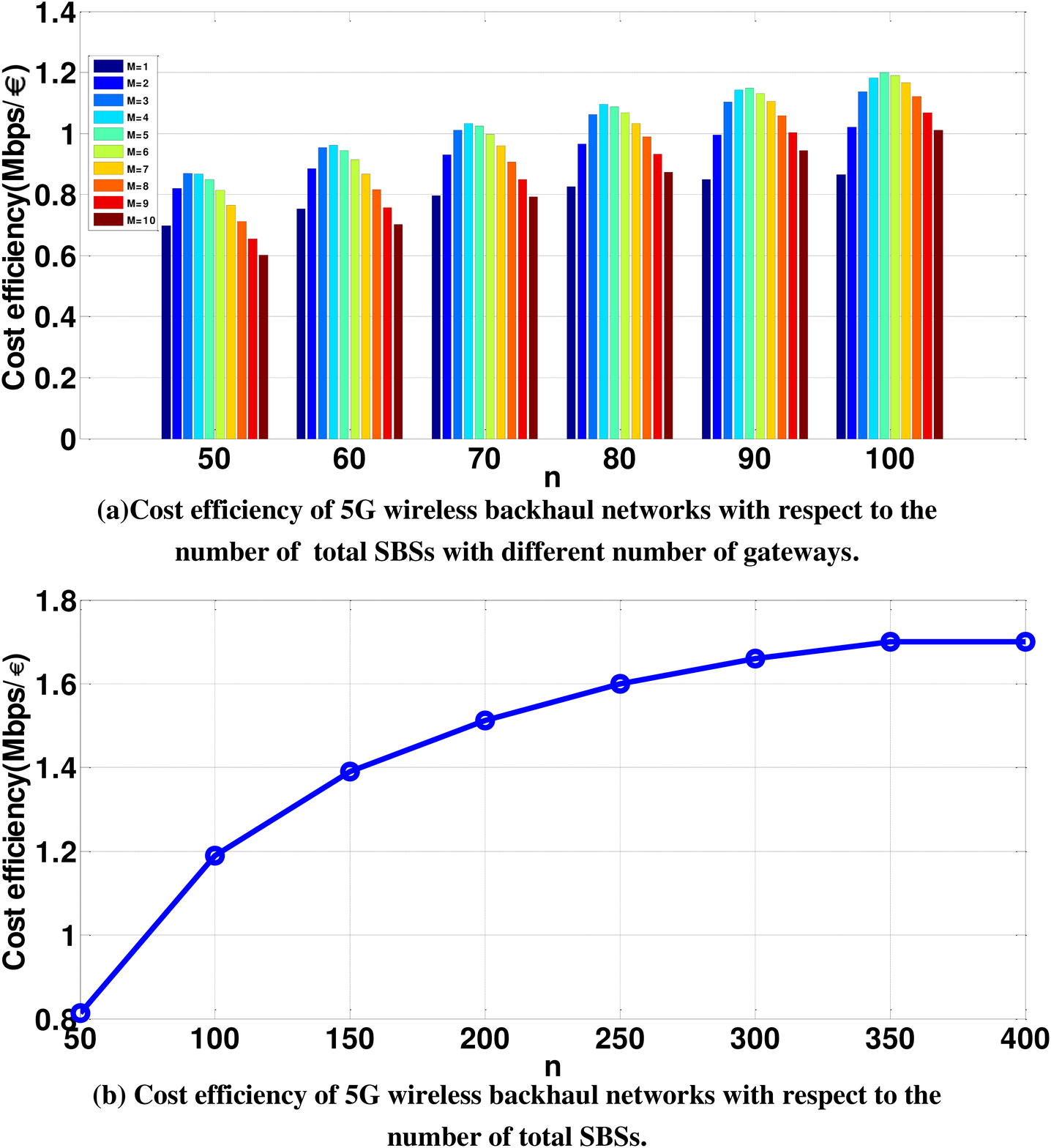}}
\caption{Cost efficiency of 5G wireless backhaul networks with respect to the number of total SBSs with different number of gateways. }
\label{fig6}
\end{figure}

Fig.~\ref{fig6} illustrates the cost efficiency of 5G wireless backhaul networks with respect to the number of total SBSs with different number of gateways. When the number of gateways is fixed in Fig.~\ref{fig6}(a), the cost efficiency of 5G wireless backhaul networks increases with the increase of the number of SBSs. When the number of total SBSs is fixed in Fig.~\ref{fig6}(a), there is a maximum cost efficiency in 5G wireless backhaul networks with different number of gateways. The reason of existing the maximal value is that the cost of 5G wireless backhaul networks increases linearly with the increase of the number of gateways, but the transport capacity of network does not always substantially increase. Moreover, the number of gateways corresponding to the maximum cost efficiency increases with the increase of the number of total SBSs in 5G wireless backhaul networks. Fig.~\ref{fig6}(b) describes the cost efficiency of 5G wireless backhaul networks with respect to the number of total SBSs when the number of gateways is equal to 6. The results of Fig.~\ref{fig6}(b) imply that the cost efficiency approach to an upper limit, i.e., 1.7008 Mbps/{\texteuro}, when the density of SBSs is larger than 350.

\begin{figure}[!t]
\centerline{\includegraphics[width=9cm, draft=false]{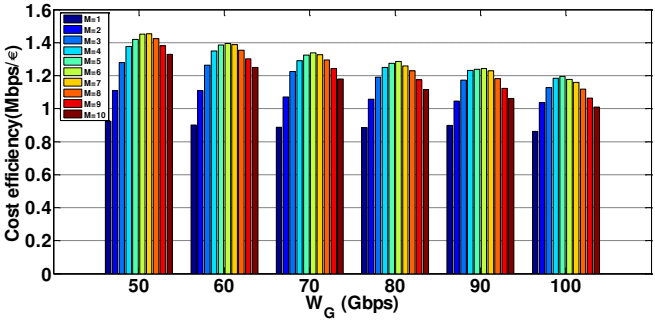}}
\caption{Cost efficiency of 5G wireless backhaul networks with respect to the gateway maximum transmission rate considering different number of gateways. }
\label{fig7}
\end{figure}

Fig.~\ref{fig7} shows the Cost efficiency of 5G wireless backhaul networks with respect to the gateway maximum transmission rate considering different number of gateways. When the number of gateways is fixed, the cost efficiency of 5G wireless backhaul networks decreases with the increase of the gateway maximum transmission rate. When the gateway maximum transmission rate is fixed, the cost efficiency of 5G wireless backhaul networks first increases with the increase of the number of gateways and then decreases with the increase of the number of gateways after the cost efficiency reaches the given maximum.

\begin{figure}[!t]
\centerline{\includegraphics[width=9cm, draft=false]{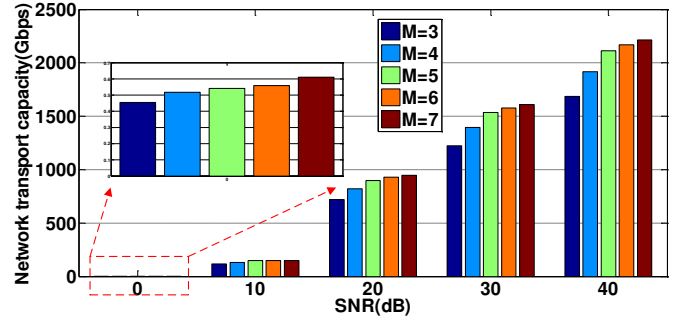}}
\caption{Network transport capacity of 5G wireless backhaul networks with respect to the SNR over wireless channels and different number of gateways. }
\label{fig8}
\end{figure}

Fig.~\ref{fig8} shows the network transport capacity of 5G wireless backhaul networks with respect to the SNR over wireless channels and different number of gateways. When the number of gateways is fixed, the network transport capacity of 5G wireless backhaul networks increases with the increase of SNR values over wireless channels. When the SNR value is fixed, the network transport capacity of 5G wireless backhaul networks increases with the increase of number of gateways.

\begin{figure}[!t]
\centerline{\includegraphics[width=9cm, draft=false]{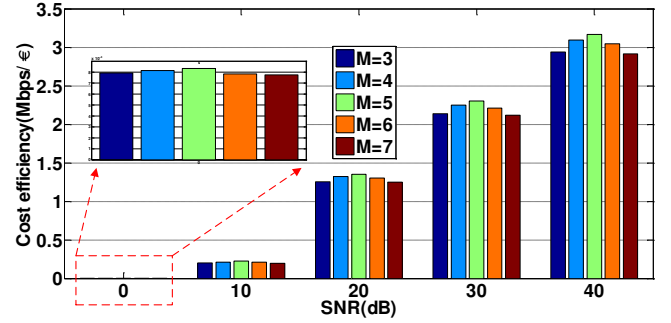}}
\caption{Cost efficiency of 5G wireless backhaul networks with respect to the SNR over wireless channels and different number of gateways. }
\label{fig9}
\end{figure}

Fig.~\ref{fig9} describes the cost efficiency of 5G wireless backhaul networks with respect to the SNR over wireless channels and different number of gateways. When the number of gateways is fixed, the cost efficiency of 5G wireless backhaul networks increases with the increase of the SNR values over wireless channels. When the SNR value is fixed over wireless channels, there is a maximum cost efficiency in 5G wireless backhaul networks with different number of gateways. Moreover, the optimal number of gateways corresponding to the maximum cost efficiency is 5.

\begin{figure}[!t]
\centerline{\includegraphics[width=9cm, draft=false]{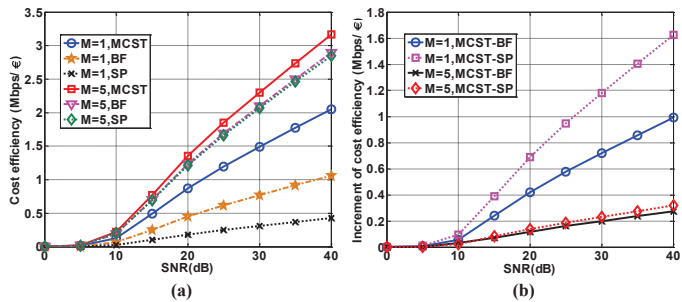}}
\caption{Cost efficiency of 5G wireless backhaul networks with respect to the SNR over wireless channels and different number of gateways. }
\label{fig10}
\end{figure}

To compare the increment of cost efficiency improved by the MCST algorithm, the increment of cost efficiency among the MCST, Bellman-Ford (BF) \cite{R32} and shortest path (SP) \cite{R27,R28} algorithms with respect to the SNR values considering different numbers of gateways is illustrated in Fig.~\ref{fig10}(b), in which the increment between the MCST and BF algorithms is labelled as "MCST-BF" and the increment between the MCST and SP algorithms is labelled as "MCST-SP". When the number of gateway is configured as 1, the maximum increment of cost efficiency are 94\% and 381\% between the MCST and BF algorithms and between the MCST and SP algorithms in Fig.~\ref{fig10}(b), respectively. When the number of gateways is 5, the maximum increment of cost efficiency are 10\% and 13\% between the MCST and BF algorithms and between the MCST and SP algorithms in Fig.~\ref{fig10}(b), respectively.

\begin{figure}[!t]
\centerline{\includegraphics[width=9cm, draft=false]{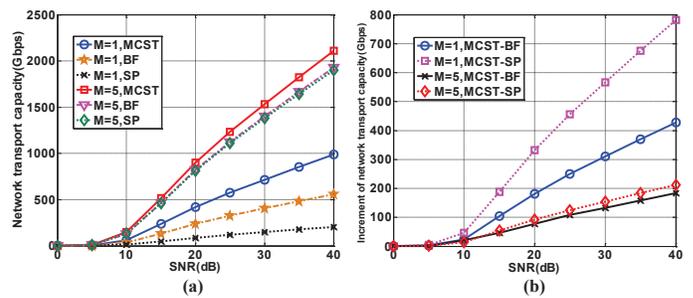}}
\caption{Network transport capacity of 5G wireless backhaul networks with respect to the SNR over wireless channels and different number of gateways. }
\label{fig11}
\end{figure}

 Fig.11 describes the network transport capacity of 5G wireless backhaul networks with respect to the MCST, BF and SP algorithms considering different SNR values. Based on the results in Fig.~\ref{fig11}(a), the network transport capacity of MCST algorithm is always larger than that of BF and SP algorithms in 5G wireless backhaul networks. The reason of this result is that the MCST algorithm can dynamically change the routes as the wireless link states are changed. As a consequence, the wireless channel capacity of the routes scheduled by the MCST algorithm is larger than or equal to the wireless channel capacity of the routes scheduled by the BF and SP algorithms. When the number of gateway is configured as 1, the maximum network transport capacity of the proposed MCST algorithm is improved by 77\% and 380\% compared with the BF and SP algorithms in Fig.~\ref{fig11}(b), respectively. When the number of gateway is configured as 5, the maximum network transport capacity of the proposed MCST algorithm is improved by 10\% and 13\% compared with the BF and SP algorithms in Fig.~\ref{fig11}(b), respectively.

 Fig.6 and Fig. 7 analyze the impact of number and locations of gateways implemented by Algorithm 2 on the cost efficiency of 5G wireless backhaul networks. Considering the number of SBSs and the gateway maximum transmission rate, the optimal number and locations of gateways can be selected by Algorithm 2 which can achieve the maximum cost efficiency of 5G wireless backhaul network in a long time scale. Fig. 9 and Fig. 10 investigate the impact of wireless channel conditions implemented by Algorithm 3 on the cost efficiency of 5G wireless backhaul networks. When the optimal number of gateways is fixed by Algorithm 2, the wireless backhaul route can be selected by Algorithm 3 based on the wireless channel conditions, i.e., SNR values in a short time scale. Moreover, the increment between the Algorithm 3 and conventional BF and SP algorithms is illustrated in Fig. 10. Based on the results of Fig. 6, Fig. 7, Fig. 9 and Fig.10, the cost efficiency of 5G wireless backhaul networks can be improved by Algorithm 2 and Algorithm 3.

\section{Conclusions}
In this paper, a two-scale cost efficiency optimization algorithm is proposed for 5G wireless backhaul networks. In the long time scale, the number and positions of gateways are optimized by the LTO algorithm. In the short time scale, the transport capacity of network is optimized by the MCST algorithm. Numerical results show that there is an optimal number of gateways for the maximum cost efficiency of 5G wireless backhaul networks and the MCST algorithm can significantly improve the cost efficiency of 5G wireless backhaul networks. Our results provide useful guideline for the deployment and optimization of 5G wireless backhaul networks.

\end{document}